\documentclass[smallextended]{svjour3}  

\usepackage{amsmath}
\usepackage{url}
\usepackage{amssymb}
\usepackage{amsfonts}
\usepackage{algorithmic}
\usepackage{graphicx}
\usepackage{textcomp}
\usepackage{xcolor}
\usepackage{hyperref}
\usepackage{subcaption}
\usepackage{balance}
\usepackage{booktabs} 
\usepackage{makecell} 
\usepackage{multirow}
\usepackage{pifont}
\usepackage{threeparttable}
\usepackage{stfloats}
\usepackage{pdflscape} 
\usepackage{graphicx}
\usepackage{tcolorbox} 
\usepackage{float}
\usepackage{adjustbox}
\usepackage{orcidlink}
\usepackage{natbib}
\usepackage{longtable}
\usepackage{array}
\usepackage{tabularx}
\usepackage[table]{xcolor}

\usepackage[dvipsnames]{xcolor}

\newcommand{\cmark}{\ding{51}}%
\newcommand{\xmark}{\ding{55}}%

\providecommand{\bmhead}[1]{\par\medskip\noindent\textbf{#1}\par\smallskip}

\def\BibTeX{{\rm B\kern-.05em{\sc i\kern-.025em b}\kern-.08em
    T\kern-.1667em\lower.7ex\hbox{E}\kern-.125emX}}



\begin{document}

\makeatletter

\let\cline\cmidrule

\makeatother

%\begin{frontmatter}

\title{Cataloguing Hugging Face Models to Software Engineering Activities: Automation and Findings}
\titlerunning{Cataloguing Hugging Face Models to Software Engineering Activities}

% \author[upc]{Alexandra González}
% \ead{alexandra.gonzalez.alvarez@upc.edu}

% \author[upc]{Xavier Franch}
% \ead{xavier.franch@upc.edu}

% \author[smu]{David Lo}
% \ead{davidlo@smu.edu.sg}

% \author[upc]{Silverio Martínez-Fernández}
% \ead{silverio.martinez@upc.edu}

% \address[upc]{Universitat Politècnica de Catalunya, Barcelona, Spain}
% \address[smu]{Singapore Management University, Singapore}

\author{
Alexandra González\orcidlink{0009-0003-7634-0343}\textsuperscript{1} \and
Xavier Franch\orcidlink{0000-0001-9733-8830}\textsuperscript{1} \and
David Lo\orcidlink{0000-0002-4367-7201}\textsuperscript{2} \and
Silverio Martínez-Fernández\orcidlink{0000-0001-9928-133X}\textsuperscript{1}
}

\institute{
\textsuperscript{1} Universitat Politècnica de Catalunya - BarcelonaTech (UPC), Barcelona, Spain \\
\email{\{alexandra.gonzalez.alvarez, xavier.franch, silverio.martinez\}@upc.edu}
\and
\textsuperscript{2} Singapore Management University, Singapore \\
\email{davidlo@smu.edu.sg}
}

% \date{}

\maketitle

\begin{abstract} 
\textit{Context}: 
Open-source Pre-Trained Models (PTMs) provide extensive resources for various Machine Learning (ML) tasks, yet these resources lack a classification tailored to Software Engineering (SE). 
\textit{Objective}: 
To address this gap, we derive a taxonomy encompassing 146 SE tasks and apply an SE-oriented classification to PTMs in a popular open-source ML repository, Hugging Face (HF). 
\textit{Method}: 
Our repository mining study followed a five-phase pipeline: (i) identification of SE tasks from the literature; (ii) collection of PTM data from the HF API, including model card descriptions and metadata, and the abstracts of the associated arXiv papers; (iii) text processing to ensure consistency; (iv) a two-phase validation of SE relevance, involving humans and LLM assistance, supported by five pilot studies with human annotators and a generalization test; and (v) data analysis. This process yielded a curated catalogue of \textcolor{black}{5,027} SE PTMs.
\textit{Results}: 
We find that most SE PTMs target \textit{code generation} and \textit{coding}, emphasizing implementation over early or late development stages. In terms of ML tasks, \textit{text generation} dominates within SE PTMs. Notably, the number of SE PTMs has increased markedly since 2023 Q2, while \textcolor{black}{benchmarking} remains limited: only \textcolor{black}{5.55\% of all SE PTMs \textcolor{black}{in our catalogue} provide evaluation results}.
\textit{Conclusions}: 
Our catalogue reveals documentation and transparency gaps, \textcolor{black}{highlights SE activity imbalances, and supports automated SE scenarios, such as the selection of suitable PTMs.}
\end{abstract}

% \begin{keyword}
% \end{keyword}

%\end{frontmatter}

\section{Introduction}
The exponential growth of open-source \textcolor{black}{registries} such as Hugging Face (HF) \citep{huggingface} has \textcolor{black}{broadened} access to Machine Learning (ML) assets \citep{zhao2024empirical}. In particular, HF is distinguished by its extensive and fast-growing collection of Pre-Trained Models (PTMs) \citep{han2021pre,ait2024hfcommunity}, compared to other platforms \citep{10.1145/3569934,10.1145/3643991.3644907}. In \textcolor{black}{2025} alone, HF saw an average of \textcolor{black}{3,044} new PTMs published per day, based on our analysis of data retrieved through the platform’s public API. Moreover, it is estimated that a new repository is created every 15 seconds \citep{huggingfaceclemHugging}, reflecting trends observed in prior studies that reported rapid increases in repository creation and model contributions \citep{10.1145/3643991.3644898}. Despite this growth, the current categorization of HF resources remains general-purpose, often targeting ML domains as broad as computer vision or natural language processing. This generic taxonomy fails to address the specific needs of the Software Engineering (SE) community, whose requirements differ significantly from those of other fields \citep{barenkamp2020applications}. Importantly, SE goes beyond mere coding, encompassing a set of activities structured around the well-known Software Development Life Cycle (SDLC) \citep{ruparelia2010software,sommerville2015}. This cycle comprises five main activities: requirements engineering, software design, software implementation, software quality assurance, and software maintenance, thereby covering the entire process from initial specification to long-term system evolution.

For this reason, SE researchers \citep{giray2021software} and practitioners \citep{tan2024challenges,zhao2024empirical} face challenges in identifying HF's ML assets that suit their tasks, while meeting project-specific constraints (e.g., licensing requirements or minimum performance scores). Common issues include missing attributes, discrepancies between reported and actual performance, and risks related to privacy or unethical model behaviour \citep{jiang2023empirical}. The absence of an SE-specific categorization in HF \citep{10.1145/3661167.3661215} sets a significant barrier, by limiting the efficient application of ML in SE. This gap has severe consequences for SE practitioners and researchers. In the absence of tailored filtering mechanisms, users must manually sift through thousands of PTMs to sample those relevant for their tasks, an effort that is both time-consuming and error-prone, ultimately slowing progress and reducing reproducibility \citep{8804457}. Although search engines can facilitate retrieval to some extent, they rely on heterogeneous metadata and lack an SE-specific focus, which favours popular PTMs over those better suited to SE tasks. Moreover, the lack of a task-aligned taxonomy prevents the SE community from systematically identifying relevant resources, evaluating progress, and enabling full automation in the multi-agent environment. 

Given the dynamic and evolving nature of both ML and SE \citep{10752650}, there is a clear need for a flexible and domain-specific cataloguing system. Such a system would support the effective sampling, selection, and integration of PTMs tailored to the full SDLC. The ability to identify suitable alternatives remains essential for sustainable and adaptable development practices. Moreover, open-source registries such as HF are founded on the principle of choice, offering visibility and reuse opportunities beyond the most popular PTMs. A structured taxonomy can therefore enhance discoverability, guiding users towards PTMs that better fit specific SE needs. 

Prior studies have taken important steps toward organizing ML resources for SE, but each presents limitations that leave a significant gap unaddressed. \citet{10.1145/3695988} proposed a taxonomy by categorizing LLM-related papers into SE tasks and SDLC activities through a systematic literature review. However, their approach is grounded in static academic sources and lacks connection to actual open-source registries like HF. \citet{10.1145/3661167.3661215} extracted SE-relevant information from HF model cards using a semi-automated method. Yet, their validation was limited to only three PTMs, and their taxonomy does not align with the SDLC, reducing its applicability. Similarly, \citet{yang2024ecosystemlargelanguagemodels} curated code-focused LLMs from HF, providing useful insights into that specific subset, but their scope excludes broader SE tasks and lacks an SDLC perspective. \citet{wang2025software} surveyed benchmarks for code LLMs and agents around the SDLC. However, their work does not address the classification of other types of PTMs.

In this work, we address these limitations by introducing an automated cataloguing framework for SE tasks across the SDLC. Our approach systematically maps PTMs from HF to a taxonomy of 146 SE tasks, built on prior research and established guidelines. We analyze multiple \textcolor{black}{textual} sources, including model card descriptions, metadata, and abstracts from associated arXiv papers. The pipeline begins with the extraction of candidate task tokens from textual descriptions and applies several processing steps to improve precision. These include outlier detection to remove inconsistent entries, similarity-based deduplication to eliminate near-identical variants, and LLM support to validate that the assigned activity reflects the model’s described purpose. Our main contributions are as follows: 

\begin{itemize}
    \item[(a)] An automated cataloguing framework that retrieves, processes, and validates PTMs from the HF API, classifying each according to the full SDLC.
    \item[(b)] An analysis of SE PTMs, covering their distribution across SE activities and tasks, cross-activity reuse, and their key attributes, including temporal evolution, base models, benchmarks, training datasets, and licenses.
    \item[(c)] A validated dataset of \textcolor{black}{5,027} SE PTMs as of \textcolor{black}{May 31, 2026}.
\end{itemize}

By providing an SDLC-aligned catalogue of PTMs, our work supports researchers and practitioners by automating the identification and selection of the most appropriate PTMs for specific tasks. This automation reduces manual effort, improves consistency in benchmarking, and accelerates the adoption of ML-driven tools within SE workflows, ultimately fostering more effective and targeted automation across the SDLC. In addition, our analysis uncovered that the number of SE PTMs in HF has grown significantly in the past few years, being heavily concentrated on software implementation.

This article is structured as follows. 
% Section~\ref{sec:related_work} reviews related work on PTMs and their application in SE. 
Section~\ref{sec:methodology} describes the methodology used to catalogue PTMs hosted on HF. Section~\ref{sec:results} presents the main results and analysis. Section~\ref{sec:discussions_limitations} discusses the findings, \textcolor{black}{reviews related work,} and explores implications. Section~\ref{sec:threats} identifies threats to validity; and Section~\ref{sec:conclusion} concludes the article and suggests future research directions.

\section{Methodology} \label{sec:methodology}
In this section, we define the goal and objectives of our study, together with the research questions that guided our investigation. We also provide the rationale behind our approach and describe the different phases of the study. 

\subsection{Research Objectives}\label{subsec:rqs}
Following the Goal Question Metric (GQM) template \citep{caldiera1994goa}, our goal is to \textbf{\textit{analyze} HF PTMs \textit{for the purpose of} their automatic classification \textit{with respect to} their application to SE activities and tasks \textit{from the point of view of} software engineers \textit{in the context of} model selection}.

This goal is structured around three RQs. The first focuses on key challenges in the automatic cataloguing of PTMs, providing essential information for the subsequent questions:
\begin{itemize}
    \item \textbf{RQ1}: What challenges arise when automatically cataloguing SE PTMs in HF? 
    
    - \textbf{RQ1.1}: What is the current documentation status of SE PTMs?
    
    - \textbf{RQ1.2}: What are the key findings regarding the correct cataloguing of SE PTMs?
\end{itemize}

Next, we examine the role of PTMs in SE, focusing on their applications across the entire SDLC:
\begin{itemize}
    \item \textbf{RQ2}: How are HF PTMs utilized for supporting SE tasks?
    
    - \textbf{RQ2.1}: What SE activities and tasks are covered by PTMs?

    - \textbf{RQ2.2}: How are cross-activity PTMs applied across multiple SE activities?
    
    - \textbf{RQ2.3}: How are ML tasks related to SE activities?
\end{itemize}

Lastly, we explore selected characteristics of SE PTMs, including temporal evolution, their underlying base model, associated benchmark scores, training datasets, and licensing information, to provide an in-depth understanding of their properties.
\begin{itemize}
    \item \textbf{RQ3}: What attributes characterize SE PTMs in HF?
    
        - \textbf{RQ3.1}: How has the number of SE PTMs evolved over time? 

        - \textbf{RQ3.2}: Which base model \textcolor{black}{families are} most frequently used for SE over time?
        
        - \textbf{RQ3.3}: What patterns emerge from the benchmark scores of SE PTMs?
        
        - \textbf{RQ3.4}: Which training datasets are used to develop SE PTMs?
        
        - \textbf{RQ3.5}: What types of licenses are applied to SE PTMs?
\end{itemize}

By addressing these RQs, we aim to understand the current landscape of SE PTMs, thereby characterizing the broader SE PTM ecosystem. Furthermore, by building a database that consolidates these PTMs and their metadata, we seek to support SE researchers and practitioners in discovering suitable PTMs for their projects.

\subsection{Research Method}
\begin{figure}[!h]
    \centering    
    \includegraphics[width=1\linewidth]{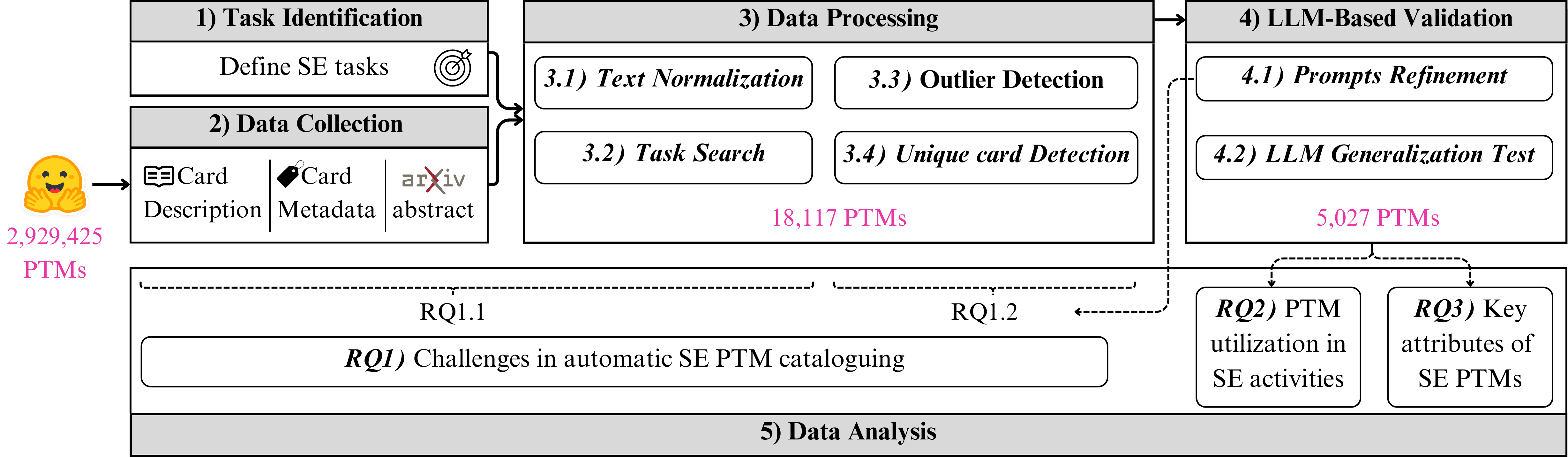}
    \caption{Pipeline to catalogue and analyze SE PTMs in HF \textcolor{black}{(May 2026 data)}} 
    \label{fig:pipeline}
\end{figure}

Following the ACM/SIGSOFT Empirical Standards \citep{githubEmpiricalStandardsdocsstandardsMaster} for mining repositories, we designed a study to classify PTMs hosted on HF using the pipeline illustrated in Figure~\ref{fig:pipeline}. The process comprises five main phases. In the first phase (\textit{Task Identification}), we identified 146 SE tasks from the literature to serve as the basis for the cataloguing. In the second phase (\textit{Data Collection}), we extracted information from each PTM available on the HF platform, including its model card description, metadata, and related arXiv abstracts. The third phase (\textit{Data Processing}) involved several steps to normalize the collected textual data, detect SE task matches, remove outliers, and identify unique model cards. In the fourth phase (\textit{LLM-Based Validation}), we used Gemini to refine the catalogue by retaining only PTMs closely aligned to SE. Finally, in the fifth phase (\textit{Data Analysis}), we analyzed the data gathered throughout the pipeline to answer each of the RQs. 

\textcolor{black}{The resulting pipeline was applied to a detailed snapshot of HF collected in March 2025, which provided an overview of the quality and quantity of PTMs available in the platform, including the completeness of their model cards. To account for the rapidly evolving nature of the platform, we subsequently collected a more recent snapshot in May 2026 and report the analysis of SE PTMs identified in this updated snapshot.}
% The following paragraphs describe each phase in detail.

This method supports reproducibility, allowing anyone with access to the replication package \citep{replication_package} to validate the results. Additionally, the pipeline is designed to be easily updated, so researchers can reapply it to classify new PTMs added to the platform using the same method. The inclusion criteria for identifying SE PTMs in HF are summarized in Table \ref{tab:inclusion_criteria}.

\begin{table}[!h]
    \centering
    \caption{Inclusion criteria to identify SE PTMs in HF}
    \label{tab:inclusion_criteria}
    \begin{tabular}{l} 
    \toprule
    (1) PTMs must be available in HF. \\
    (2) PTMs must have a valid card description, metadata, or an abstract \\ from an associated paper on arXiv. \\
    (3) The card description, metadata, or abstract must specify an SE task. \\
    (4) Gemini must confirm relevance to an SE activity. \\
    \bottomrule
    \end{tabular} 
\end{table}

\subsubsection{Task Identification}\label{subsubsec:task_identification} To identify SE tasks applicable to the five activities of the SDLC, we built on prior literature and extended it to form a broader taxonomy. \textcolor{black}{We started from the work of \citet{10.1145/3695988}, which, to our knowledge, offers the broadest SDLC-aligned coverage of SE tasks. It comprises 88 SE tasks organized into six SE activities: requirements engineering, software design, software development, software quality assurance, software maintenance, and software management.}

\textcolor{black}{After discussion among the authors, we adapted this baseline to the scope of our study by (i) renaming software development to software implementation, since \textit{development} often denotes the entire SDLC, whereas \textit{implementation} restricts the activity to building the software, consistent with \citet{sommerville2015}; (ii) excluding software management, which cuts across the lifecycle rather than constituting a phase of it; (iii) explicitly incorporating software architecture tasks into software design, as they were scarcely represented despite their role in this activity; and (iv) removing \textit{agile story point estimation}, which concerns project planning rather than software construction, as well as the generic \textit{others} task in software implementation and maintenance.}

\textcolor{black}{As this starting point derives from a literature review on LLMs up to January 2024, it covers only tasks that had already been studied in that context. To capture tasks across the entire SDLC regardless of their level of automation, we extended it using the Software Engineering Body of Knowledge (SWEBOK) \citep{washizaki2024swebok}, mapping its knowledge areas to our activities: requirements engineering (Chapter 1), software design (Chapters 2 and 3), software implementation (Chapter 4), software quality assurance (Chapters 5, 12, and 13), and software maintenance (Chapter 7). From the key concepts and topics of each chapter, we retained those that were not already represented in the baseline taxonomy. This added 16 tasks in requirements engineering, 14 in software design, 5 in software implementation, 22 in software quality assurance, and 6 in software maintenance. The resulting taxonomy comprises a total of \textbf{146 tasks} (see Appendix \ref{sec:se_tasks}).}

\subsubsection{Data Collection} We accessed the HF API \citep{huggingfaceHuggingFace} to retrieve all PTMs available as of \textcolor{black}{May 31, 2026}. For each identifier, we retrieved\textcolor{black}{, when present,} (i) its \textit{model card description}, a markdown file documenting the model's characteristics, intended uses, and evaluation \citep{mitchell2019model}; (ii) the associated \textit{card metadata}, specified in a YAML block (e.g., license, tags, language) \citep{huggingfaceModelCardsMetadata}, which has been shown to help users find appropriate PTMs \citep{suryani2025model}; and (iii) the \textit{abstract} of a linked arXiv paper \citep{huggingfaceModelCardsPaper}, taking advantage of the cross-platform linking between HF and arXiv \citep{suryani2024exploration}. Any of these components may be missing for a given PTM. At that time, there were \textcolor{black}{\textbf{2,929,425 PTMs}} available in HF.

\subsubsection{Data Processing}
The third phase of the methodology involves four sequential steps to process the collected data, ranging from \textcolor{black}{(i)} text normalization to standardize matches during \textcolor{black}{(ii)} task \textcolor{black}{search}, \textcolor{black}{(iii)} identifying outliers and \textcolor{black}{(iv)} unique instances, ensuring that no PTMs are duplicated or counted twice. 

% \begin{figure}[!h]
%     \centering
%     \includegraphics[width=0.8\linewidth]{pipeline_processing.pdf}
%     \caption{\textcolor{black}{Data processing workflow (May 2026 data)}}
%     \label{fig:data_processing_pipeline}
% \end{figure}

\paragraph{\textbf{3.1) Text Normalization}} We preprocessed the list of SE tasks and the collected data to ensure accurate SE task detection across heterogeneous textual sources. Model cards, metadata, and abstracts in HF exhibit diverse writing styles and formats, as terms may appear in uppercase or inflected forms (e.g., ``Code Summarization", ``code summarisation"  or ``code summarizing"). To handle this variability, we applied tokenization, lowercasing, and lemmatization. %\textit{Tokenization} splits the text into smaller units, such as words, which simplifies the subsequent analysis, while \textit{lowercasing} eliminates inconsistencies arising from variations in capitalization. Additionally, \textit{lemmatization} reduces words to their base forms, allowing different inflections to be treated as a single entity. 
Together, these steps standardize the textual descriptions, ensuring greater catalogue accuracy.

\paragraph{\textbf{3.2) Task Search}} As we focused on SE-relevant resources, we searched for the name of each SE task within the prepared data. When a task consisted of multiple words (e.g., ``code generation"), we required all tokens to appear together and in the correct order to be considered a valid match. Moreover, since the search was strictly token-based, task names embedded within other tokens were systematically excluded.

\paragraph{\textbf{3.3) Outlier Detection}} We applied several techniques to refine the available resources and ensure that only the most relevant and accurate ones were included in our catalogue. Our primary focus was on detecting positive outliers, i.e., SE tasks whose frequency was abnormally high compared to the rest, as these cases typically signal false positives caused by lexical ambiguities rather than actual SE relevance. We found that \textit{debugging}, \textit{logging}, and \textit{coding} had notably higher occurrences, together accounting for \textcolor{black}{almost 90\%} of task occurrences. For \textit{debugging}, some entries contained \texttt{debug:None} in the metadata, which triggered a match in the search, though it should not have. Similarly, for \textit{logging}, the lemma ``log" often appeared in code snippets or reporting tables, such as those involving logging functions (e.g., \texttt{import logging}) or entries like  \texttt{train log}, which were not relevant to the task. For \textit{coding}, we considered a model to be for coding only if it included the tag ``code" in the metadata. To ensure robustness of the filtering, we revisited the raw data for these PTMs without lemmatizing, as we wanted to understand the context in which the lemmas appeared, applying the exclusion criteria in Table \ref{tab:exclusion_criteria}. We then checked for any occurrences that did not meet our exclusion criteria.

\begin{table}[!h]
    \centering
    \caption{Exclusion criteria for the outlier detection filtering}
    \label{tab:exclusion_criteria}
    \begin{tabular}{lll}
        \toprule
        \textbf{Activity} & \textbf{Task} & \textbf{Description} \\
        \midrule
        Software & Debugging & (i) `debug' after `.', `-', `\_', `/', `\textless', `\textgreater' \\
        Maintenance & & (ii) `debug' before `-', `\_', `/', `:', `\textless', `\textgreater' \\
        & & (iii) `debug' followed by `` or ' \\
        & & (iv) `` or ' followed by `debug' \\
        \midrule 
        Software & Logging & (i) `log' followed by an alphabetic \\ 
        Maintenance & & character, but not `g' \\
        & & (ii) `log' between a period and a letter \\
        & & (iii) `log' after `.', `-', `\_', `/', `(', `)' \\
        & & (iv) `log' before `-', `\_', `/', `:', `(', `)' \\
        & & (v) ``import logging" \\
        & & (vi) ``no log" \\
        & & (vii) ``training log" \\
        & & (viii) ``train log" \\
        & & (ix) ``testing log" \\
        & & (x) ``test log" \\
        \midrule
        Software & Coding & (i) absence of the `code' tag \\
        Implementation & \\
        \bottomrule
    \end{tabular}
\end{table}

\paragraph{\textbf{3.4) Unique Card Detection}} During preliminary analysis, we observed that many model cards exhibit high textual similarity. For instance, the model card of \texttt{deepseek-ai/DeepSeek-V3-Base} \citep{huggingfaceDeepseekaiDeepSeekV3BaseHugging} is nearly identical to those of other PTMs on the platform, such as \texttt{deepseek-ai/Deep\\Seek-V3} \citep{huggingfaceDeepseekaiDeepSeekV3Hugging} and \texttt{Conexis/DeepSeek-R1-NextN-Channel-\\INT8}~\citep{huggingfaceConexisDeepSeekR1NextNChannelINT8Hugging}. In particular, duplicate and near-duplicate PTMs often emerge because users re-upload base models with little or no modification \citep{horwitz2025charting}.
To address this observation, we applied a \textbf{duplicate detection} step using \textit{term frequency–inverse document frequency (TF-IDF)} and \textit{cosine similarity}. We set a 0.99 similarity threshold, which is high enough to detect near-identical cards while allowing truly distinct versions to remain separate because their documentation differs sufficiently to fall below this threshold. Without this step, we risk double-counting PTMs that have minimal documentation differences but are otherwise the same model. This procedure resulted in \textcolor{black}{\textbf{18,117 PTMs}, as of May 2026}, with the full list and their high-similarity matches available in the replication package \citep{replication_package}.

\subsubsection{LLM-Based Validation}
\begin{figure}[!h]
    \centering
    \includegraphics[width=1\linewidth]{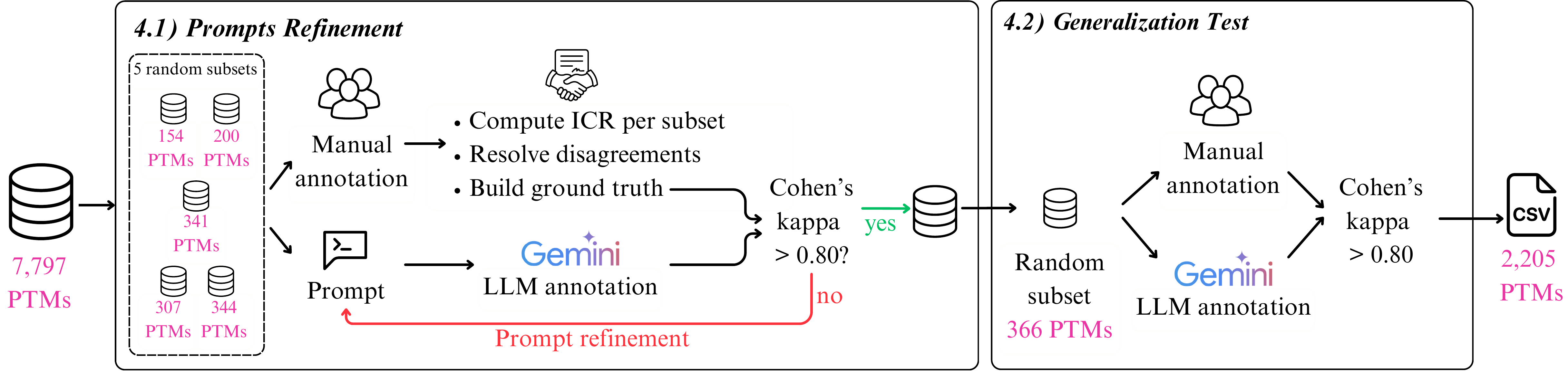}
    \caption{Process of using an LLM to validate the catalogue \textcolor{black}{(March 2025 data)}}
    \label{fig:pipeline_LLM}
\end{figure}

In the fourth phase of the methodology, we used an LLM to validate the cataloguing and ensure that only SE-relevant resources \textcolor{black}{were retained}. To improve reliability, we conducted a two-phase validation procedure, as illustrated in Figure~\ref{fig:pipeline_LLM}. \textcolor{black}{This process was carefully developed using the first snapshot (March 2025 data), and the resulting validated prompts were subsequently applied to the second one (May 2026 data).} All prompt iterations and manually annotated subsets are available in the replication package \citep{replication_package}.

\paragraph{\textbf{4.1) Prompts Refinement}}\label{subsubsec:pilot} We employed Gemini 2.0 Flash, selected due to its strong benchmark performance \citep{rahman2025comparative}, to assess the relevance of PTMs to SE activities. \textcolor{black}{When using this LLM, we followed the steps of the PRIMES framework proposed by \cite{de2024experiences}. Specifically, we followed their guidelines when creating and piloting the prompts, and validating the output.}

We conducted five pilot tests (one per SE activity) to validate the LLM. For this reason, we selected  \textbf{five random subsets} and ensured representation across all their tasks. The sample sizes were determined using a sample size calculator, applying a 95\% confidence level and a 5\% margin of error to ensure statistical validity \citep{qualtricsSampleSize}. Table \ref{tab:pilot_subset} presents the unique PTMs per SE activity and the corresponding pilot sample sizes. Note that a given PTM can be associated with more than one SE activity.

\begin{table}[!htb]
    \centering
    \caption{PTMs per SE activity and pilot sample sizes \textcolor{black}{with ICR subset (March 2025 data)}}
    \label{tab:pilot_subset}
    \begin{tabular}{lrrr}
        \toprule
        \textbf{SE Activity} & \textbf{Unique PTMs} & \textbf{Pilot Sample} & \textcolor{black}{\textbf{ICR Subset}} \\
        \midrule
        Requirements Engineering & 255 & 154 & \textcolor{black}{15} \\
        Software Design & 415 & 200 & \textcolor{black}{20} \\
        Software Implementation & 2,993 & 341 & \textcolor{black}{34} \\
        Software Quality Assurance & 1,512 & 307 & \textcolor{black}{31} \\
        Software Maintenance & 3,255 & 344 & \textcolor{black}{35} \\
        \bottomrule
    \end{tabular}
\end{table}

\textcolor{black}{To establish the ground truth, human validation was performed following a four-step process.} \textcolor{black}{First,} the first author manually annotated all pilot samples for each SE activity. In addition to providing a binary judgment (yes/no) on whether a PTM was relevant to a given SE activity, the annotator also documented the underlying reasoning behind the decision. \textcolor{black}{Second,} to ensure reliability, 10\% of the pilot samples were independently annotated by the second and fourth authors, both knowledgeable in the domain, drawing on established intercoder reliability (ICR) practices in qualitative research \citep{o2020intercoder}. \textcolor{black}{Third,} to consolidate our annotation guidelines and understanding, the three annotators participated in a joint discussion session focused on the cases where disagreement had occurred. This discussion revealed that some PTMs are inherently challenging to classify due to their broad or ambiguous descriptions. For example, \texttt{mradermacher/Regex-AI-Llama-3.2-1B-GGUF} \citep{huggingfaceMradermacherRegexAILlama321BGGUFHugging} was initially assigned to software implementation based on the ``code generation" tag included by the author in the card metadata. While the description primarily refers to RegEx generation, this tag was an explicit link to the software implementation activity, and we ultimately retained the label. \textcolor{black}{Fourth,} for the 10\% of the pilot samples, we applied majority voting to resolve any discrepancies. The resulting decisions, combined with the initial annotations from the first author, are referred to as the \textbf{ground truth}.

We then prompted the LLM in a zero-shot setting to provide both a binary judgment and a rationale for each PTM in the pilot set. \textcolor{black}{The prompts followed a structured format comprising five components, as summarized in the box below.} We used the default temperature setting of the model to balance determinism and variability in responses, ensuring consistent yet informative outputs. \textcolor{black}{All prompt iterations developed during the refinement process are available in the replication package \citep{replication_package}.}

\begin{tcolorbox}[title=Prompt Structure, colback=gray!5!white, colframe=gray!75!black, sharp corners, boxrule=0.8pt]
The prompt followed a structured format comprising:
\begin{enumerate}
    \item \textbf{Objective}: Assess the PTM's applicability to the target activity.
    \item \textbf{Background context}: Provide a brief overview of the SDLC and the specific SE activity being evaluated.
    \item \textbf{Task description}: Instructions on how to review and assess the PTM for the target SE activity.
    \item \textbf{Criteria}: Explicit rules for determining relevance to the target SE activity.
    \item \textbf{Example output}: Format for binary judgment and rationale.
\end{enumerate}
\end{tcolorbox}

The model’s performance was evaluated by comparing its responses with the human annotations using Cohen’s kappa \citep{sim2005kappa}. This process was repeated iteratively, refining the prompt until we reached almost perfect agreement, as seen in Table \ref{tab:pilot_kappa}. This process was conducted for each of the five SE activities. In total, 5 prompt refinements were required for requirements engineering, 12 for software design, 4 for software implementation, 9 for software quality assurance, and 8 for software maintenance. The refinements focused on providing examples of SE tasks per activity, explicitly instructing the use of the criteria when making judgments, and avoiding superficial or ambiguous associations with the activity based on the information provided.

\begin{table}[!htb]
    \centering
    \caption{Cohen’s kappa for \textcolor{black}{human-LLM agreement (pilot, March 2025)}} 
    \label{tab:pilot_kappa}
    \begin{tabular}{lcl}
        \toprule
        \textbf{SE Activity} & \textbf{Cohen’s Kappa} & \textbf{Agreement} \\ 
        \midrule
        Requirements Engineering & 1.00 & Perfect \\
        Software Design & 0.85 & Almost perfect \\
        Software Implementation & 0.85 & Almost perfect \\
        Software Quality Assurance& 0.81 & Almost perfect \\
        Software Maintenance & 0.82 & Almost perfect \\
        \bottomrule
    \end{tabular}
\end{table}

\paragraph{\textbf{4.2) Generalization Test}} To evaluate the LLM’s ability to generalize beyond the initial piloting subset and avoid prompt overfitting, we evaluated it on a separate, manually labeled evaluation subset that was not involved in any piloting activities. This set consisted of 366 randomly sampled PTMs across all SE activities, a \textcolor{black}{statistically representative sample (95\% confidence, 5\% margin)} for our population. The ground truth for this subset was created using the same rigorous process as before: the full subset was annotated by the first author, and 10\% of it was independently labeled. Then, we instructed the LLM with the five prompts and computed Cohen’s kappa to measure agreement between human decisions and the LLM’s responses on this subset, observing an almost perfect level of agreement ($\kappa = 0.88$). This result demonstrates a high degree of generalizability and provides confidence to use the LLM to label the remaining samples. 
Lastly, we applied the five prompts to all resources, resulting in \textbf{\textcolor{black}{5,027} PTMs (May 2026 snapshot)} meeting our inclusion criteria. \textcolor{black}{Given the time elapsed between prompt validation and the second snapshot, we updated the underlying LLM to Gemini 2.5 Flash and verified that the prompts remained valid.}

\subsubsection{Data Analysis}
The final step of the methodology involves analyzing all the data gathered, from the initial data dump from HF to the final set of SE PTMs. 

\paragraph{\textbf{RQ1 Analysis.}} RQ1 focuses on identifying the main challenges in automatically cataloguing SE PTMs in HF.
To address RQ1.1, we relied on the \textcolor{black}{March 2025 snapshot} collected from HF, consisting of 1,533,973 PTMs. Within this data, we examined the completeness and quality of its documentation. We computed descriptive statistics on the availability of model cards and associated arXiv abstracts, and whether they included any SE task match.

To study contextual misclassifications (RQ1.2), we analyzed the frequency distribution of SE task tokens across all PTMs and applied outlier detection to isolate abnormally frequent lemmas. The contextual meaning of each outlier was manually reviewed to determine whether occurrences were relevant or false positives. This combination of quantitative and qualitative analysis provided insights into the documentation quality and ambiguity of SE PTM cataloguing.

\paragraph{\textbf{RQ2 Analysis.}}
For RQ2, we \textcolor{black}{used the May 2026 snapshot and} examined the utilization of PTMs across SE activities, the extent to which PTMs span multiple activities, and their alignment with underlying ML tasks. Each of the 5,027 validated PTMs was associated with one or more SE tasks from the taxonomy, enabling the characterization of their distribution throughout the SDLC (RQ2.1). 

The cross-activity analysis was conducted by examining co-occurrence patterns between activities to identify PTMs associated with multiple ones (RQ2.2). These patterns were represented as a co-occurrence matrix that allowed examination of relationships among activities without assuming exclusivity. 

Furthermore, to study the connection between SE activities and ML tasks, we exploited the task-based organization currently used in HF, which categorizes models according to their underlying ML task types, and computed their joint distributions with the SE catalogue (RQ2.3).

\paragraph{\textbf{RQ3 Analysis.}}
The analyses under RQ3 characterized SE PTMs through temporal, structural, and legal dimensions of SE PTMs.
For temporal evolution (RQ3.1), we parsed the creation date field of each PTM, normalized timestamps into quarterly intervals, and computed cumulative growth curves to describe the evolution of SE-related models over time. 

To analyze base-model reuse (RQ3.2), we extracted the \texttt{base\_model} attribute, introduced by HF in November 2023, and aggregated PTMs by base model and creation date to assess dependency patterns among variants of the same base model (e.g.,  fine-tuned or derivative models).

For benchmark characterization (RQ3.3), we parsed the \texttt{model-index} section of each card to identify evaluation results. Since naming conventions across cards are heterogeneous, we applied a rule-based normalization to standardize benchmarks and metrics (e.g., benchmark variants such as \textit{MMLU} and \textit{Measuring Massive Multitask Language Understanding}, or metrics such as \textit{accuracy\_norm} and \textit{normalized\_accuracy}). This analysis is motivated by the growing interest in SE benchmarks in recent years: for SE LLMs, the number of benchmarks increased from 30 in 2022 to 81 in 2024 \citep{hu2025assessing}. 

To analyze training datasets (RQ3.4), we processed the \texttt{datasets} field in the cards to find patterns in dataset reuse. We mapped dataset usage trends and their alignment with SE activities.

Finally, for licensing (RQ3.5), we extracted license information to assess reuse conditions. Licensing considerations have become increasingly important in the development of PTMs, as recent studies highlight the need for license awareness to mitigate copyright risks \citep{lo2023trustworthy}. Accordingly, we checked how permissive, copyleft, and restrictive licenses are used across SE activities.

\section{Results}\label{sec:results}
In this section, we present the findings from our study, organized according to the three RQs introduced in Section~\ref{subsec:rqs}. 

\subsection{What challenges arise when automatically cataloguing SE PTMs in HF? (RQ1)}
\textcolor{black}{Using the detailed snapshot of March 2025,} we examine the main challenges encountered when automatically cataloguing SE PTMs in HF. We first analyze the current documentation status of SE PTMs to assess the completeness and reliability of available model information. Next, we discuss the key findings that emerged during the classification process. %, including the handling of duplicated model cards and the disambiguation of SE task mentions across different contexts.

\subsubsection{What is the current documentation status of SE PTMs?}
\textcolor{black}{To assess the documentation of HF PTMs and the extent to which it provides SE information, we examine their model cards and associated arXiv papers.}
As summarized in Table \ref{tab:cards_summary}, 68.52\% of the PTMs have a non-empty model card, and particularly 7.32\% mention an SE task. This aligns with prior findings that model information can be missing or inaccurate due to the self-reporting nature of model metrics \citep{jiang2023empirical}.

\begin{table}[!h]
    \centering
    \caption{Availability and content of model cards and arXiv abstracts for PTMs in HF \textcolor{black}{(March 2025 data)}}
    \label{tab:cards_summary}
    \begin{tabular}{llrr}
        \toprule
        \textbf{Source} & \textbf{Category} & \textbf{PTMs} & \textbf{Proportion} \\
        \midrule
        Model Card & Not available & 478,125 & 31.17\% \\
         & Available but empty & 4,774 & 0.31\% \\
         & Available, no SE tasks & 938,815 & 61.20\% \\
         & Available, with SE tasks & 112,259 & 7.32\% \\
         \midrule
        arXiv Abstract  & Not available or no SE tasks & 1,531,643 & 99.85\% \\
         & Available, with SE tasks & 2,330 & 0.15\% \\
        \midrule
        \multicolumn{2}{l}{Total unique PTMs with SE tasks} & 113,558 & 7.40\% \\
        \multicolumn{2}{l}{Total unique PTMs} & 1,533,973 & 100.00\% \\
        \bottomrule
    \end{tabular}
\end{table}

To complement this, we analyzed the abstracts of the associated arXiv papers linked to all PTMs. This source revealed that only 0.15\% of all PTMs have an associated abstract that mentions SE tasks, while the remaining 99.85\% either lack an arXiv ID or do not mention any SE task.

By combining all sources, we identified a total of \textbf{113,558 unique PTMs} that could be associated with SE tasks. This total includes 1,299 additional PTMs identified exclusively through arXiv abstracts, which were not flagged as SE-relevant based on their model card description and metadata alone.

\vspace{0.1cm}
\noindent
\fcolorbox{black}{white}{
    \parbox{0.95\linewidth}{
    \textbf{Finding 1.1}: The state of HF \textcolor{black}{as of March 2025} shows that PTMs scarcely report SE tasks in model cards.
    
    \textbf{Finding 1.2}: \textcolor{black}{Based on the March 2025 snapshot,} arXiv abstracts associated with PTMs can provide valuable insights into their SE applicability.
    }
}
\vspace{0.1cm}

\subsubsection{What are the key findings regarding the correct cataloguing of SE PTMs?}
\textcolor{black}{We examine the challenges found when identifying SE PTMs, focusing on duplicated model cards and the contextual interpretation of SE task mentions.}
The application of our classification pipeline revealed that a significant number of model cards are duplicated. After identifying SE task matches and removing outliers, we found that only 52.11\% of the PTMs were actually unique based on their documentation. This duplication is primarily due to two reasons: either they correspond to different variants of the same model (e.g., optimized or fine-tuned versions), or users re-upload PTMs under their own namespace without making significant modifications to the card content. These duplications reflect the existence of an ecosystem in which PTMs depend on and interact with each other \citep{yang2024ecosystemlargelanguagemodels}. For example, we observed that \texttt{sparsh35/Meta-Llama-3.1-8B-Instruct}~\citep{huggingfaceSparsh35MetaLlama318BInstructHugging} has 290 highly similar counterparts, such as \texttt{triplee/torchtune\_8B\_full\_finet\\uned\_llama3.1\_millfield\_241223\_meta\_before\_user\_epoch2} \citep{huggingfaceTripleetorchtune_8B_full_finetuned_llama31_millfield_241223_meta_before_user_epoch2Hugging}. Similarly, the PTM \texttt{dducnguyen/newdream-sdxl-20} \citep{huggingfaceDducnguyennewdreamsdxl20Hugging} is associated with 715 near-duplicate PTMs. To mitigate this, we implemented a duplicate detection strategy whereby the earliest published model card is treated as the original, and all others with high textual similarity are labeled as similar variants of it.

During the pilot manual annotation phase, we detected that some tasks could be interpreted in different contexts. To ensure consistency, we used the LLM with explicit inclusion and exclusion criteria for each SE activity, allowing it to disambiguate cases based on context. Notable examples include: 

\begin{itemize}
    \item \textbf{Requirements engineering}: Terms such as \textit{system/software requirement} often referred to the technical prerequisites for using the PTM rather than functional requirements in an SE context. Similarly, \textit{use case generation} was frequently misidentified due to a recurring pattern in many model cards (particularly in Gemma models) where the phrase appeared as part of the ``Risks identified and mitigations'' section, without referring to requirements engineering. 
    
    \item \textbf{Software design}: Expressions like \textit{rapid prototyping}, \textit{architectural design} (used in the context of interior design or neural network architectures), and \textit{software/system design} (commonly present in phrases such as ``software designed to...'') did not reflect SE-specific practices but triggered false-positive identifications. 
    
    \item \textbf{Software implementation}: Many PTMs labeled with \textit{API inference} or \textit{data analysis} were unrelated to SE, as they were often linked to generative tasks like Stable Diffusion or generic ML applications. 
    
    \item \textbf{Software quality assurance}: The term \textit{verification} appeared in contexts such as speaker verification, fact verification, or self-verification, which are unrelated to software verification. Similarly,  \textit{cybersecurity} was often mentioned in relation to red teaming assessments. 
    
    \item \textbf{Software maintenance}: We found cases where \textit{code revision} referred to dataset hashes rather than code updates, and \textit{sentiment analysis} was used in other contexts (e.g., general natural language processing tasks).
\end{itemize}

\vspace{0.1cm}
\noindent
\fcolorbox{black}{white}{
    \parbox{0.95\linewidth}{
    \textbf{Finding 1.3}: Correct classification of SE PTMs requires addressing duplicated model cards and contextual misinterpretations of SE tasks.
    }
}
\vspace{0.1cm}

\subsection{How are HF PTMs utilized for supporting SE tasks? (RQ2)}
To answer RQ2, we \textcolor{black}{use the snapshot of May 2026 and} analyze how PTMs are applied across different SE activities and tasks. We first examine the coverage of SE tasks to identify which areas of the SDLC are supported by PTMs. Next, we explore PTMs that span multiple SE activities to understand cross-activity reuse, and finally, we investigate the relationships between SE activities and ML tasks to highlight patterns of ML support across the \textcolor{black}{SDLC}.

\subsubsection{What SE activities and tasks are covered by PTMs?}
\textcolor{black}{We first characterize the coverage of SE activities and tasks by examining how the identified PTMs are distributed across the SDLC.}
Our taxonomy includes 146 SE tasks across the five SE activities of the SDLC \textcolor{black}{(see the complete list in Appendix~\ref{sec:se_tasks})}, but only a subset of them was represented among the PTMs identified in HF. After completing all phases of our pipeline, we found that \textcolor{black}{74} SE tasks were covered in the final validated set of \textcolor{black}{5,027} SE PTMs. \textcolor{black}{The remaining 72 SE tasks had no associated PTM as of May 2026. These gaps span all five SE activities. Examples include \textit{requirements elicitation} and \textit{requirements scrubbing} in requirements engineering; \textit{architecture style} and \textit{software specification synthesis} in software design; \textit{API inference} and \textit{API documentation smells} in software implementation; \textit{fault localization} and \textit{test prediction} in software quality assurance; and \textit{code revision} and \textit{vulnerability repair} in software maintenance.} 

\begin{figure}
    \centering
    \includegraphics[width=1\linewidth]{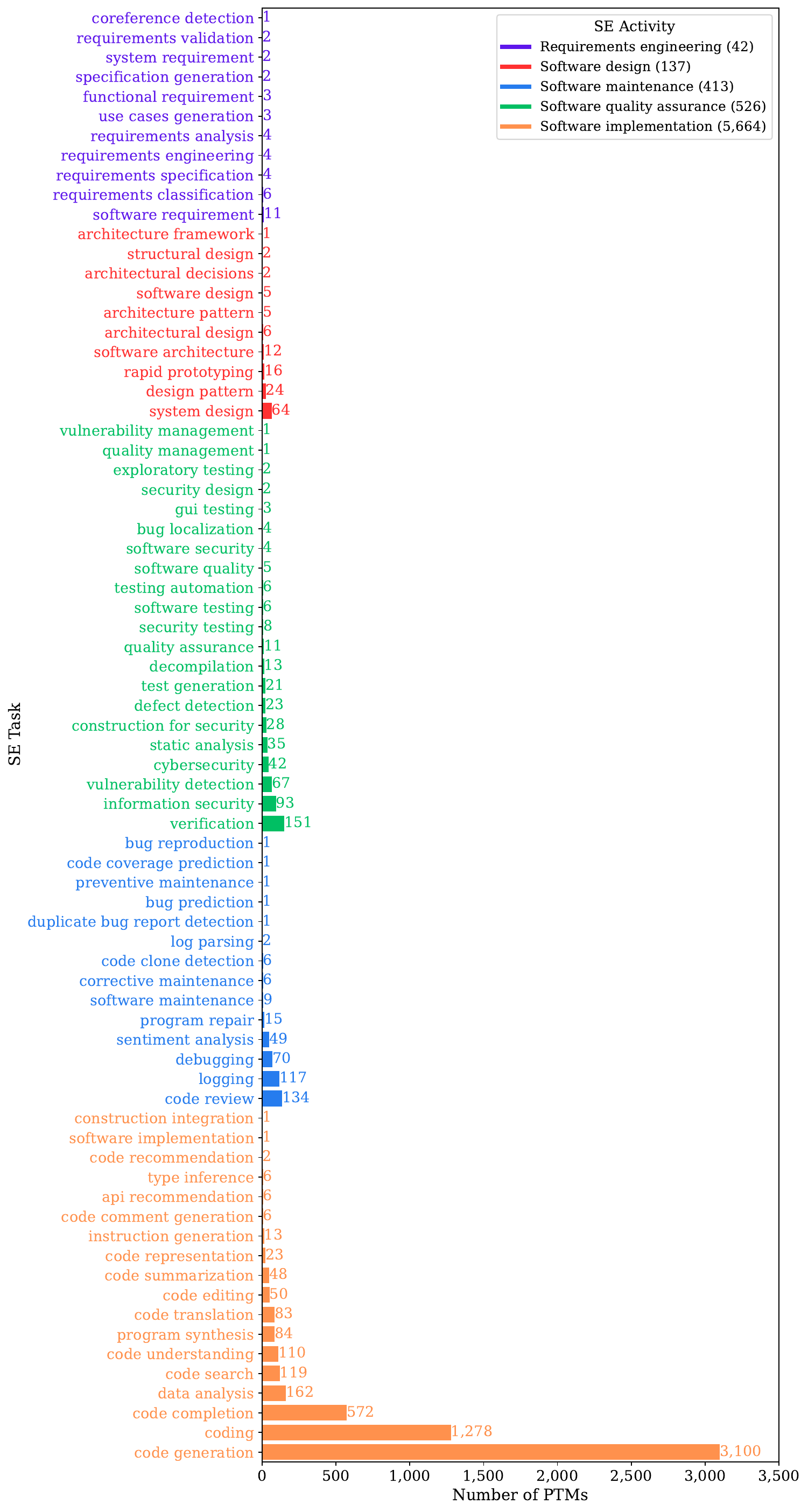}
    \caption{PTMs associated with each SE task and SE activity}
    \label{fig:mdls_barchart}
\end{figure}

Figure \ref{fig:mdls_barchart} shows the distribution of PTMs across tasks \textcolor{black}{with at least one PTM}, with each bar color-coded to reflect its corresponding SE activity. \textcolor{black}{Note that a given PTM may be associated with multiple SE tasks (e.g., \textit{code summarization} and \textit{debugging}) and, consequently, with multiple SE activities (e.g., software implementation and software maintenance). Therefore, the total number of PTMs represented in the figure exceeds the aforementioned 5,027 unique PTMs. For further insights into cross-activity support, refer to Section~\ref{subsubsec:cross-activity}.}

A clear pattern emerges: \textit{code generation} and \textit{coding} are by far the most represented SE tasks. Notably, PTMs designed for \textit{code generation} outnumber those for general \textit{coding} support by more than a twofold margin. This highlights the strong focus on text-to-code and code-to-code applications \textcolor{black}{in the current state of the art of HF PTMs}.

From an activity-level perspective, the data reveal that software implementation is the dominant SE activity in terms of PTM coverage. In contrast, early-stage activities such as requirements engineering and software design are less represented, with only a small fraction of PTMs targeting them. Later phases of the SDLC, including software maintenance and quality assurance, demonstrate moderate PTM adoption. Tasks in these phases, such as \textcolor{black}{\textit{verification}} and \textcolor{black}{\textit{code review}}, are relatively well-supported, indicating a recognition of the value of automation in ensuring software quality. 

% \vspace{0.1cm}
\noindent
\fcolorbox{black}{white}{
    \parbox{0.95\linewidth}{
     \textbf{Finding 2.1}: \textit{Code generation} and \textit{coding} are the most covered SE tasks, \textcolor{black}{as of May 2026}.
    
     \textbf{Finding 2.2}: \textcolor{black}{In the same snapshot}, software implementation-related PTMs outnumber those for early and late SDLC stages.
    }
}
\vspace{0.1cm}

\subsubsection{How are cross-activity PTMs applied across multiple SE activities?}\label{subsubsec:cross-activity}
\begin{figure}[!h]
    \centering
    \includegraphics[width=1\linewidth]{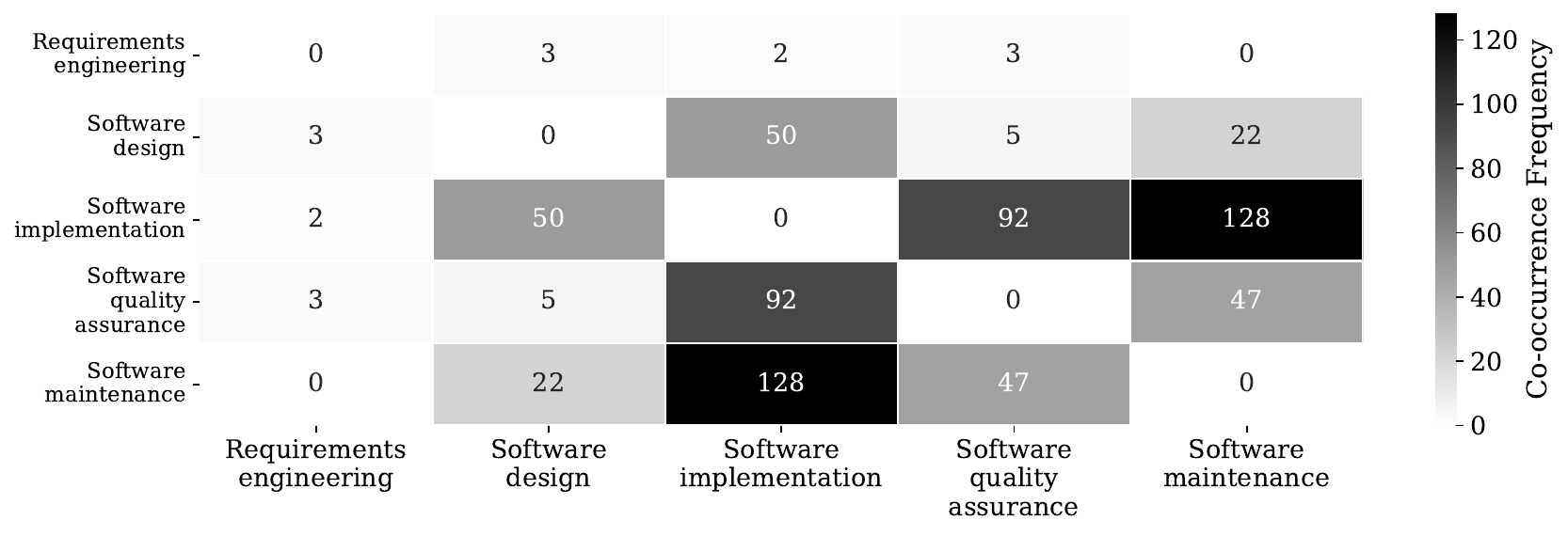}
    \caption{\textcolor{black}{Pairwise co-occurrence of SE activities across cross-activity PTMs}}
    \label{fig:cross-activity}
\end{figure}

PTMs can be related to more than one SE task, potentially spanning different SE activities (i.e., a PTM can be simultaneously classified into multiple SE activities). To better understand this phenomenon, we selected PTMs associated with more than one distinct SE activity and represented their pair-wise relationships in Figure \ref{fig:cross-activity}. The strongest co-occurrence is observed between software implementation and software maintenance, followed by links \textcolor{black}{between software implementation} and software quality assurance. 

Although the absolute number of shared PTMs is small relative to the overall corpus of \textcolor{black}{5,027} SE models, the overlaps are substantial within certain activities. For instance, \textcolor{black}{among PTMs associated with multiple SE activities, 50 out of 80 co-occurrences involving software design (62.5\%) also involve software implementation.} This suggests that certain PTMs are particularly well-suited for reuse across mid- and late-stage activities in the SDLC, whereas connections to requirements engineering remain relatively sparse. 

\vspace{0.1cm}
\noindent
\fcolorbox{black}{white}{
    \parbox{0.95\linewidth}{
     \textbf{Finding 2.3}: \textcolor{black}{As of May 2026,} shared PTMs are few in absolute terms but represent a substantial proportion within certain activities, \textcolor{black}{especially those involving software implementation}. 
    }
}
\vspace{0.1cm}

\subsubsection{How are ML tasks related to SE activities?}
\begin{figure}[!h]
    \centering
    \includegraphics[width=1\linewidth]{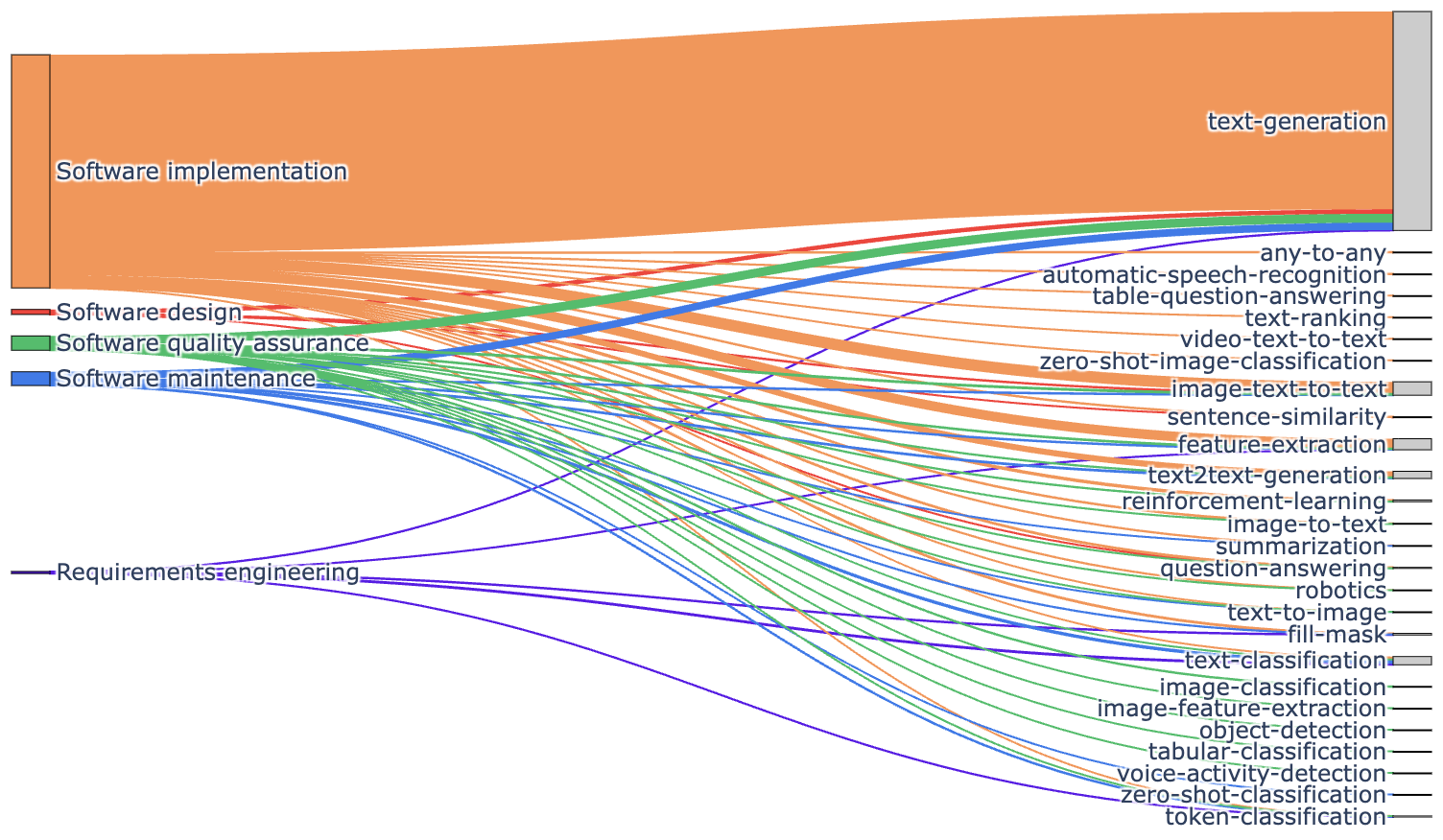}
    \caption{Association of SE activities with ML tasks}
    \label{fig:mdls_sankey}
\end{figure}

\textcolor{black}{Here, we analyze the relationship between SE activities and the ML tasks assigned in HF to characterize how ML capabilities are distributed across the SDLC. This} relationship is illustrated in Figure \ref{fig:mdls_sankey}, with the flow indicating the number of PTMs. ML tasks correspond to the current classification used in HF, i.e., a label assigned to each PTM. The figure highlights the prominence of the ML task \textit{text generation}, which is associated with the highest number of PTMs, reflecting its relevance in supporting code-related tasks. From the SE perspective, this task is most commonly linked to software implementation. However, it extends across the other SE activities, indicating its versatility. 

ML support is uneven: \textcolor{black}{early phases (requirements engineering and software design) demonstrate few connections to diverse ML tasks, while late phases (software quality assurance and maintenance) demonstrate connections to multiple ones, but notably fewer than software implementation. Examples of supported ML tasks include \textit{text generation}, \textit{text classification}, and \textit{text2text generation}, which are useful in the context of SE, such as \textit{code review} tasks.}

\vspace{0.1cm}
\noindent
\fcolorbox{black}{white}{
    \parbox{0.95\linewidth}{
     \textbf{Finding 2.4}: The ML task \textit{text generation} is the most common among SE-related PTMs \textcolor{black}{in the May 2026 snapshot}.

    \textbf{Finding 2.5}: \textcolor{black}{Early SDLC phases show limited ML task diversity, while later phases exhibit broader support, though less than implementation.}
    }
}
\vspace{0.1cm}

\subsection{What attributes characterize SE PTMs in HF? (RQ3)}
SE PTMs can be characterized by a set of attributes that provide insights into their properties, performance, and usage conditions. To examine how these attributes are documented in practice, we conducted an in-depth analysis of the information reported in HF model cards \citep{huggingfaceModelCards} \textcolor{black}{as of May 2026}. Our analysis focused on five main attributes: temporal evolution, base model, benchmark scores, training dataset, and license type.

\subsubsection{How has the number of SE PTMs evolved over time?}
\begin{figure}[!h]
    \centering
    \includegraphics[width=1\linewidth]{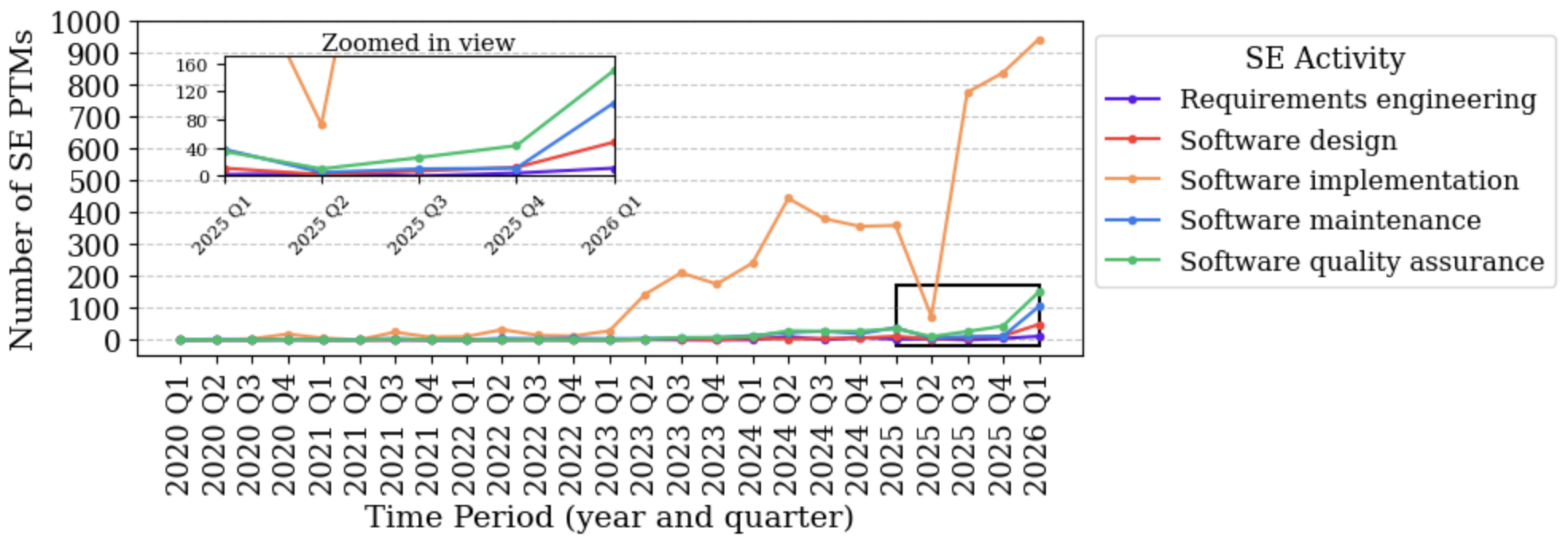}
    \caption{Number of SE PTMs created quarterly since 2020}
    \label{fig:mdls_time_inset}
\end{figure}

\textcolor{black}{We first examine the temporal evolution of SE PTMs to characterize how their development has changed over time and across SE activities.}
Figure \ref{fig:mdls_time_inset} shows growth in the creation of SE PTMs since 2020, with quarterly data revealing periods of acceleration. \textcolor{black}{Although a temporary decrease is observed in 2025 Q2 for all SE activities, PTM creation rates recover in the following quarter}. 
Notably, the number of new PTMs developed for software implementation rose from \textcolor{black}{28 in 2023 Q1 to 943 in 2026 Q1}. This upward trend intensified during the last year (from 2025 Q2 to 2025 Q4). In comparison, the current number of PTMs related to software implementation is more than \textcolor{black}{nine} times higher than that of the next most represented SE activity, software \textcolor{black}{quality assurance}. 

In parallel, we observe a steady rise in the number of PTMs associated with requirements engineering, software maintenance, software quality assurance, and software design. Although the absolute number of PTMs for these activities remains much lower than that of software implementation (as shown in Figure~\ref{fig:mdls_barchart}), the temporal trends indicate a gradual growth across all these SE activities. This suggests that, while software implementation dominates in volume, other activities are increasingly being addressed by new PTMs.

\vspace{0.1cm}
\noindent
\fcolorbox{black}{white}{
    \parbox{0.95\linewidth}{
        \textbf{Finding 3.1}: Since 2020, SE PTMs have grown steadily, with sharp acceleration observed from 2023 onward.

        \textbf{Finding 3.2}: The number of PTMs addressing SE activities continues to grow, with software implementation maintaining a strong lead.
    }
}
\vspace{0.1cm}

\subsubsection{Which base model \textcolor{black}{families are} most frequently used for SE over time?}
\begin{figure}[!h]
    \centering
    \includegraphics[width=1\linewidth]{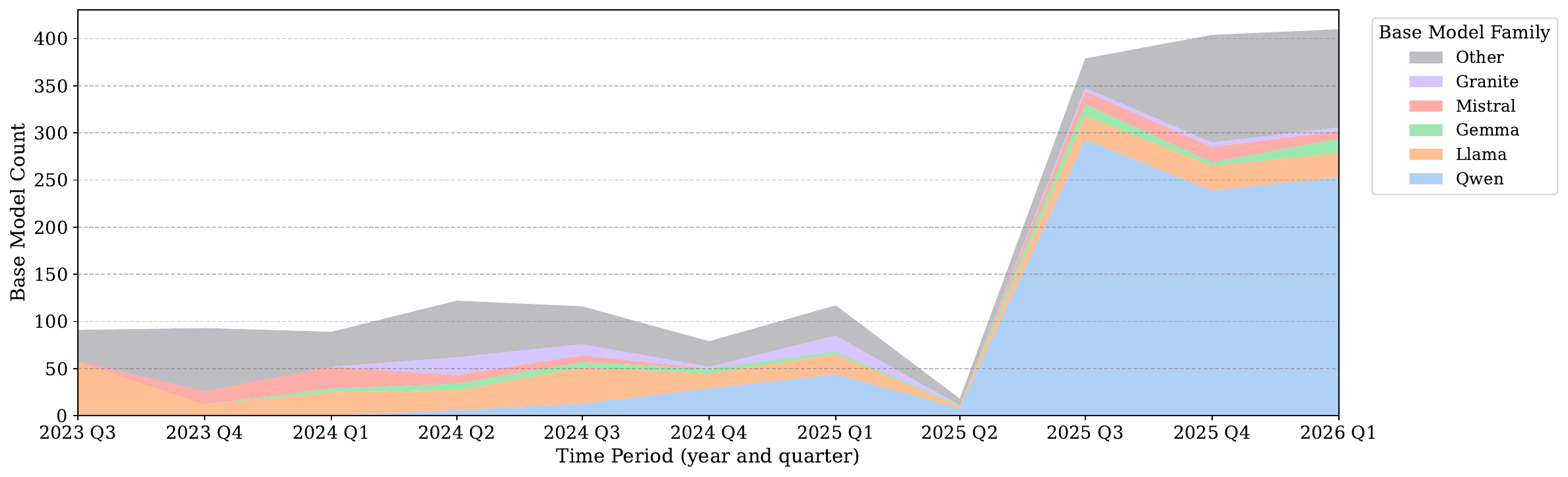} 
    \caption{\textcolor{black}{Quarterly base model reuse per family for SE PTMs since 2023 Q3}}
    \label{fig:base-mdls}
\end{figure}

We analyze the \textcolor{black}{number of SE PTMs created in each quarter that reuse different base model families, based on} the information provided in the model cards \citep{huggingfaceModelCardsBaseModel}. Figure \ref{fig:base-mdls} presents the \textcolor{black}{evolution of base model family reuse for building SE PTMs, including Qwen, Llama, Gemma, Mistral, and Granite. The analysis covers the period since HF introduced the possibility of specifying a \texttt{base\_model} attribute in model cards.}

\textcolor{black}{Consistent with the temporary decrease of PTM creation observed in Figure~\ref{fig:mdls_time_inset}, there is a sharp decline in base model reuse in 2025 Q2. Then, the reuse of base models increases sharply, approaching 410 newly created models in 2026 Q1. This rapid expansion is primarily driven by the Qwen family, which alone accounts for 253 derivative SE PTMs created in the first quarter of 2026. Other families, such as Llama, also support substantial reuse, while Gemma, Mistral and Granite contribute to PTM reuse to a lesser extent.}

\vspace{0.1cm}
\noindent
\fcolorbox{black}{white}{
     \parbox{0.95\linewidth}{
         \textbf{Finding 3.3}: \textcolor{black}{The quarterly reuse of base models for SE PTMs has increasingly relied on the Qwen family since 2025 Q2.}
     }
}
\vspace{0.1cm}

\subsubsection{What patterns emerge from the benchmark scores of SE PTMs?}
\textcolor{black}{We further examine the evaluation data reported for SE PTMs, distinguishing between evaluation coverage across SE activities and the reported score distributions among the evaluated PTMs. As of May 2026, 279 SE PTMs included evaluation information (5.55\% of all SE PTMs \textcolor{black}{in our catalogue}). Thus, the following observations are limited to this subset of SE PTMs and should not be interpreted as reflecting the overall performance of the ecosystem.}

\textcolor{black}{Table \ref{tab:mdls_benchmarks_activity} shows the evaluation coverage across SE activities. The two activities with the most PTMs, software implementation and software quality assurance, have similar evaluation coverage, at 5.89\% and 5.62\%, respectively. In contrast, the remaining activities have evaluation coverage of around 3\% or lower. For requirements engineering, \texttt{rajinikarcg/RequirementClassifier} \citep{huggingfaceRequirementClassifierDatasets} reports a score of 0 on the \textit{PROMISE NFR} benchmark, which indicates the model's inability to identify non-functional requirements. The three software design PTMs report varying scores across different benchmarks and evaluation configurations. Specifically, \texttt{Harish102005/Qwen2.5-Coder-7B\\-manim} \citep{huggingfaceHarish102005Datasets} reports 55.3\% on \textit{3blue1brown-manim}; \texttt{AlphaExaAI/ExaMind} \citep{huggingfaceAlphaExaAIDatasets} reports 94.8\% on \textit{MMLU World Religions (0-shot)} and 79.3\% on \textit{HumanEval pass@1}; and \texttt{chipcraftx-io/chipcraftx-rtlgen-7b} \citep{huggingfaceChipCraftXDatasets} reports 36.5\% on \textit{Functional Pass Rate (standalone) pass@1} and 98.7\% on \textit{Functional Pass Rate (ChipCraftX hybrid system) pass@1}. Given the very small number of evaluated PTMs in these activities and the heterogeneity of the benchmarks and configurations, these observations illustrate the limited availability of evaluation evidence for long-tail SE activities.}

\begin{table}[!htb]
    \centering
    \caption{\textcolor{black}{Evaluation coverage by SE activity}}
    \label{tab:mdls_benchmarks_activity}
    \begin{tabular}{lrrr}
    \toprule
    \textbf{SE Activity} & \textbf{Total PTMs} & \textbf{Evaluated PTMs} & \textbf{Proportion} \\
    \midrule
    Requirements engineering & 31 & 1 & 3.23\% \\
    Software design & 116 & 3 & 2.59\% \\
    Software implementation & 4,383 & 258 & 5.89\% \\
    Software quality assurance & 445 & 25 & 5.62\% \\
    Software maintenance & 371 & 12 & 3.23\% \\
    \bottomrule
    \end{tabular}
\end{table}

\begin{figure}[!h]
    \centering
    \includegraphics[width=0.95\linewidth]{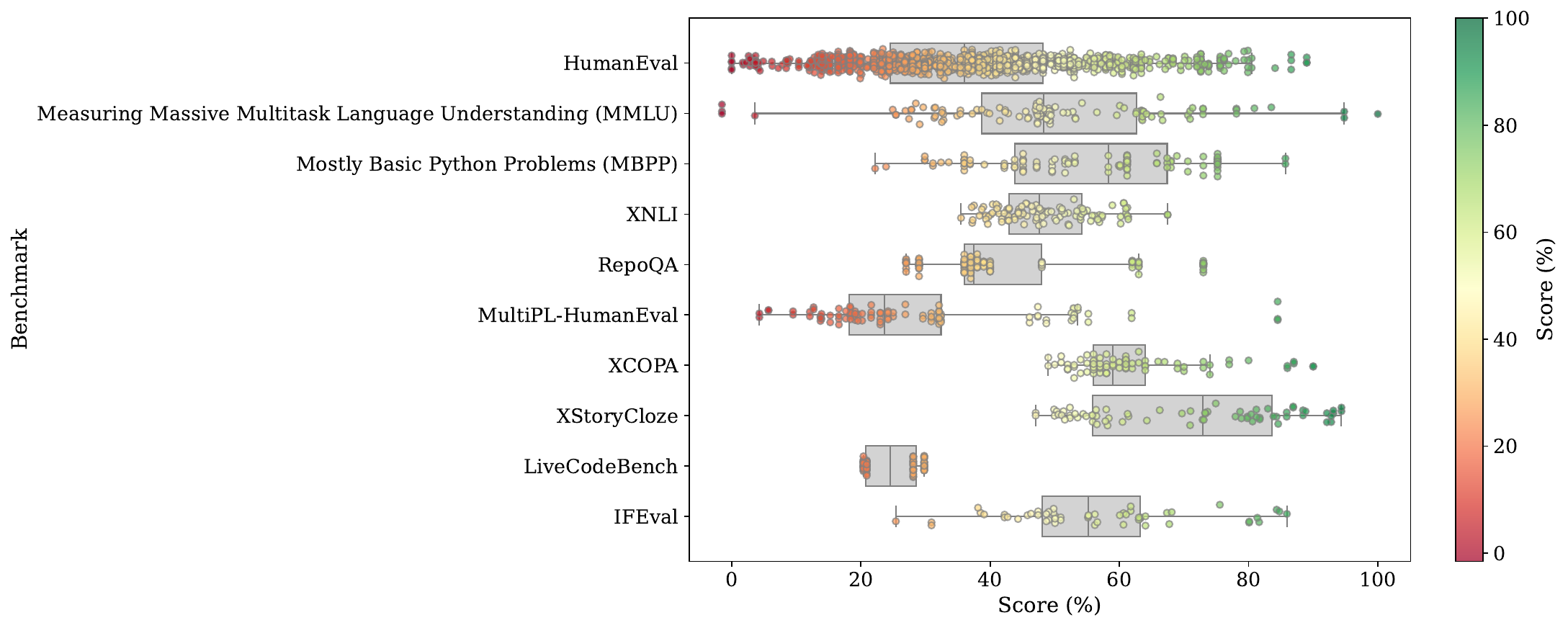}
    \caption{Score distribution for the top ten most frequently reported benchmarks in SE PTMs}
    \label{fig:mdls_benchmarks_scores}
\end{figure}

Figure \ref{fig:mdls_benchmarks_scores} illustrates the scores for the ten most frequently reported benchmarks \textcolor{black}{among the 279 evaluated SE PTMs. \textit{HumanEval}, \textit{MMLU} and MBPP dominate, accounting for the highest number of reported results. In terms of performance, PTMs evaluated on \textit{XStoryCloze}, \textit{XCOPA}, \textit{MBPP} and \textit{IFEval} achieve median scores above 50\%. \textit{HumanEval} and \textit{MMLU} are the benchmarks with the broadest score ranges, spanning from 0 to 89\% for the former and from 0 to 100\% for the latter. In contrast, \textit{LiveCodeBench} shows a comparatively narrow score range, with a median of 24.5\% and scores ranging from 20.4\% to 29.8\%.}

\vspace{0.1cm}
\noindent
\fcolorbox{black}{white}{
    \parbox{0.95\linewidth}{
         \textbf{Finding 3.4}: \textcolor{black}{In May 2026,} only \textcolor{black}{5.55}\% of SE PTMs \textcolor{black}{in our catalogue} report evaluation results, highlighting limited coverage across the ecosystem. 
        
         \textbf{Finding 3.5}: Long-tail activities, including requirements, design and maintenance, have evaluation coverage of around 3\% or lower.

         \textbf{Finding 3.6}: \textcolor{black}{\textit{HumanEval} and \textit{MMLU} exhibit the broadest score ranges among the ten most frequently reported benchmarks.}
    }
}
\vspace{0.1cm}

\subsubsection{Which training datasets are used to develop SE PTMs?}
\begin{figure}[!h]
    \centering
    \includegraphics[width=1\linewidth]{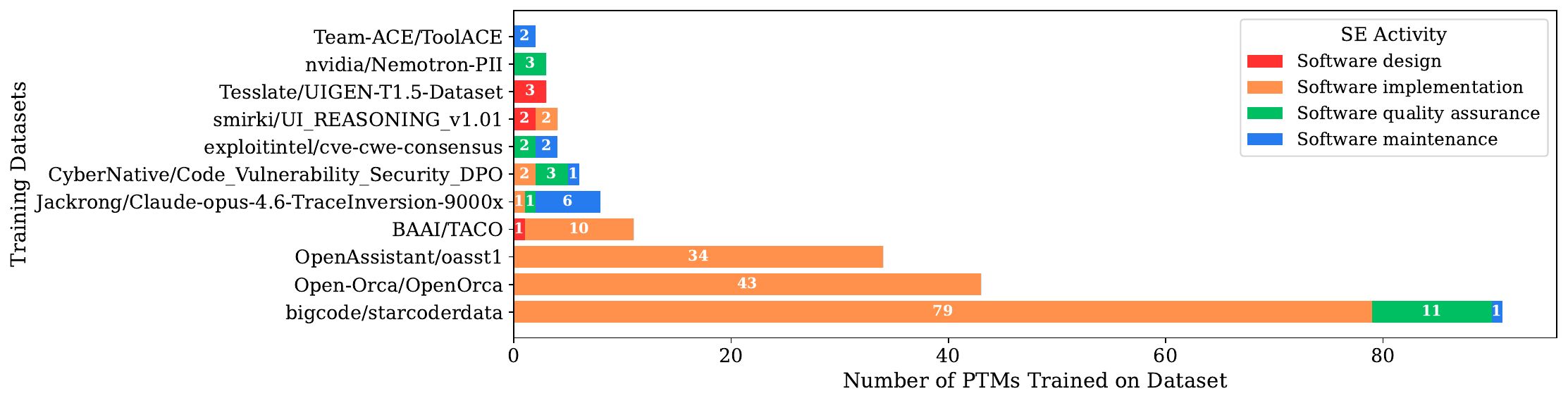}
    \caption{Top 3 training datasets per SE activity by number of PTMs trained on them}
    \label{fig:top_training_ds}
\end{figure}

\textcolor{black}{We analyze the training datasets referenced by SE PTMs to characterize patterns of dataset reuse and activity-specific use.} For the purpose of analysis, we identified the top three training datasets per SE activity based on the number of PTMs that reference each dataset. Notably, if any of these top datasets are reused across multiple SE activities, we explicitly indicate this reuse in Figure~\ref{fig:top_training_ds}. 

The results reveal clear patterns of dataset reuse across SE activities. For example, \textcolor{black}{\texttt{bigcode/starcoderdata} \citep{huggingfaceBigcodestarcoderdataDatasets}} is the most commonly reused dataset, appearing in \textcolor{black}{79} PTMs for software implementation, \textcolor{black}{11} for software quality assurance\textcolor{black}{, and 1 for software maintenance}. This reuse suggests that datasets are not always uniquely tailored to specific SE activities. 

In contrast, some datasets show activity-specific use. For instance, \textcolor{black}{\texttt{Open-\\Orca/OpenOrca} \citep{huggingfaceOpenOrcaOpenOrcaDatasets} and \texttt{OpenAssistant/oasst1} \citep{huggingfaceOpenAssistantoasst1Datasets} appear exclusively in 43 and 34 PTMs targeting software implementation, respectively}. Such patterns may reflect more deliberate curation or clearer alignment between dataset content and task requirements.

Despite this, significant gaps remain. In particular, no PTMs associated with requirements engineering specify a training dataset in their model card metadata. This absence may reflect the limited availability of standardized datasets for requirements engineering tasks or a lack of consistent reporting practices for this activity.

\vspace{0.1cm}
\noindent
\fcolorbox{black}{white}{
    \parbox{0.95\linewidth}{
         \textbf{Finding 3.7}: Some training datasets are reused across  SE activities.  
         
         \textbf{Finding 3.8}: No PTMs targeting requirements engineering report associated training datasets\textcolor{black}{, as of May 2026}.
    }
}
\vspace{0.1cm}

\subsubsection{What types of licenses are applied to SE PTMs?} 
\begin{figure}[!h]
    \centering
    \includegraphics[width=0.9\linewidth]{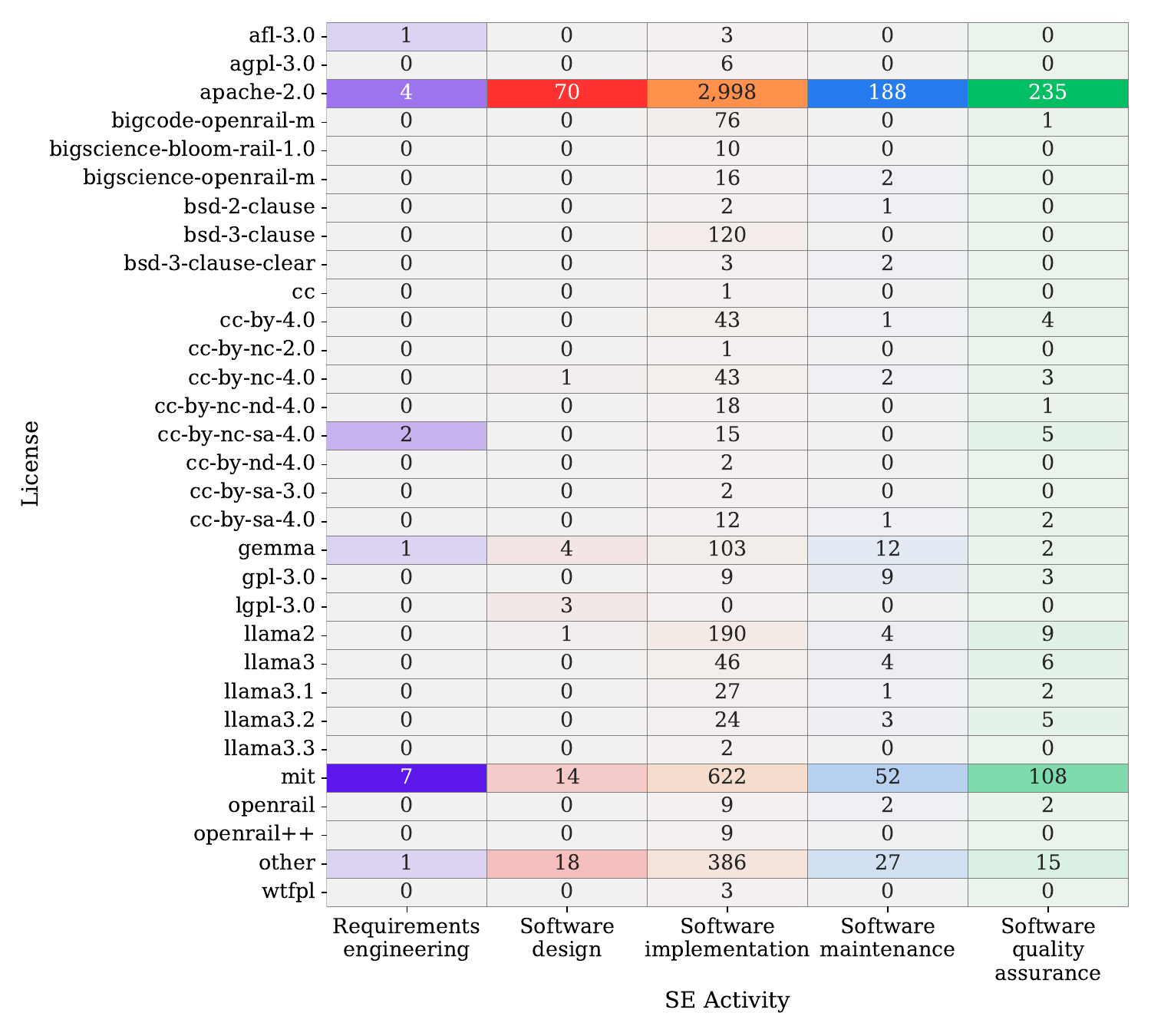}
    \caption{License distribution across SE activities}
    \label{fig:licenses}
\end{figure}

\textcolor{black}{Lastly, we study the licenses associated with SE PTMs to characterize the distribution of reuse conditions across SE activities.} As illustrated in Figure \ref{fig:licenses}, a wide variety of licenses are applied, reflecting different levels of reuse permissions and restrictions tailored to diverse needs.

Permissive licenses such as \textit{Apache 2.0} and \textit{MIT} dominate, particularly in the software implementation phase, where \textit{Apache 2.0} licenses cover a striking \textcolor{black}{2,998} PTMs and \textit{MIT} licenses cover \textcolor{black}{622} PTMs. This dominance aligns with prior works on HF, underscoring the preference for licenses that enable broad use, modification, and redistribution with minimal barriers \citep{pepe2024hugging}. 

More restrictive licenses, such as the \textit{Affero General Public License (AGPL-3.0)} and the \textit{GNU General Public License (GPL-3.0)}, are less common but notable: \textit{AGPL-3.0} \textcolor{black}{present only} in software implementation, and \textit{GPL-3.0} in implementation, maintenance \textcolor{black}{and quality assurance} phases. 

The presence of licenses such as \textit{BSD} variants (\textit{BSD-2-Clause}, \textit{BSD-3-Clause}) and \textit{Creative Commons} licenses (e.g., \textit{CC-BY-4.0}, \textit{CC-BY-NC-4.0}) illustrates additional options that balance openness with certain usage conditions, like attribution or non-commercial clauses. Overall, this diverse and skewed license distribution reflects a pragmatic balance between openness and control that shapes the adoption and reuse of SE PTMs in the HF ecosystem.

\vspace{0.1cm}
\noindent
\fcolorbox{black}{white}{
    \parbox{0.95\linewidth}{
         \textbf{Finding 3.9}: \textcolor{black}{By May 2026,} permissive licenses, especially Apache 2.0 and MIT, dominate, facilitating broad adoption and reuse.

         \textbf{Finding 3.10}: Restrictive licenses are less frequent \textcolor{black}{as of May 2026}, but notably present in specific activities.
    }
}
\vspace{0.1cm}

\section{Discussion}\label{sec:discussions_limitations}
In this paper, we present a large-scale catalogue of \textcolor{black}{5,027} SE PTMs available in HF, classified according to a fine-grained list of 146 SE tasks derived from the literature \citep{10.1145/3695988,washizaki2024swebok}. This classification provides significant insights into SE PTMs, their characteristics, and integration within the SDLC. 
This section discusses identified documentation and transparency gaps, \textcolor{black}{reflects on long tails, reviews related work}, compares our findings with prior studies and large-scale HF datasets, and highlights the practical and research implications of our generated catalogue. Lastly, we detail how our methodology can be updated for future extensions.

\subsection{Documentation gaps and transparency in SE PTMs}
A significant gap exists in the documentation of SE PTMs in HF. \textcolor{black}{As of March 2025,} over 31\% of the available PTMs lacked a model card, and only 7.32\% included a valid model card with explicit reference to an SE task. This finding reinforces earlier observations that HF model information is frequently incomplete or inaccurate due to its self-reporting nature \citep{jiang2023empirical}. Importantly, this issue is not confined to small or experimental PTMs, but also affects large projects with thousands of downloads \citep{stalnaker2025ml}. 

Prior studies confirm that incomplete documentation is a systemic problem across HF. \citet{horwitz2025charting}, in their analysis of 63,000 models from HF, found that only 8\% of them outlined license information, 13.4\% reported accuracy details, and 30.2\% specified their base model. Similarly, \citet{suryani2025model}, in a broader-scale study covering the entire HF, reported that nearly one-third of model cards include licensing information. In our SE-focused dataset, only \textcolor{black}{5.55}\% of PTMs report evaluation results, often using different metrics and benchmarks, further complicating comparative assessment. Licensing patterns among SE PTMs also reflect broader trends across the HF platform. As reported by \citet{stalnaker2025ml}, 62.5\% of all HF models are released under open-source licenses, with \textit{apache-2.0} and \textit{mit} emerging as the most common choices, consistent with our findings for SE PTMs. Notably, in our data, the ``other" license category is the third most frequent among SE PTMs, while in the overall HF corpus it ranks fifth \citep{stalnaker2025ml}. As discussed by \citet{stalnaker2025ml}, this category often appears when PTMs were uploaded before the correct license was included in HF’s official list or when model owners are unfamiliar with available licensing options.

These findings align with ecosystem-level trends identified by \citet{laufer2025anatomy}, who found that model cards within families tend to shorten and become increasingly standardized through templates or automatically generated text, mirroring our observation of model card missing information and similar documentation across PTMs. This absence of documentation reduces transparency, limits informed reuse, and creates obstacles to evaluating the suitability of PTMs for specific SE applications.

\subsection{\textcolor{black}{The long tail of SE PTMs}}
\textcolor{black}{Beyond documentation gaps, we have also observed an uneven distribution of SE PTMs in HF. In our dataset, requirements engineering and software design activities exhibit the thinnest PTM coverage, while software implementation dominates. This may reflect structural differences across SE activities and may be driven by different factors, such as research attention, industrial adoption priorities, and the complexity of automating different SE activities.}

\textcolor{black}{The research community has largely focused on implementation applications of Artificial Intelligence (AI). In the case of LLMs, \cite{10.1145/3695988} found that their adoption in SE is concentrated on software implementation (56.65\%), compared with requirements (3.9\%) and design (0.92\%). Similarly, prior studies have emphasized the limited attention given to requirements engineering in the context of LLM-based SE, with most applications focused on coding tasks \citep{fan2023large}. Likewise, there has been limited research on the application of LLMs to software design and on their benefits and drawbacks \citep{wang2026using}.}

\textcolor{black}{Although generative AI can boost SE productivity beyond code writing \citep{ebert2023generative}, the long tail may also reflect current priorities in industrial adoption. A survey of 65 software developers showed that while generative AI has a huge impact on design, implementation, testing, and documentation, lower productivity gains were perceived for early SDLC phases (i.e., planning and requirements analysis)\citep{gurgul2026state}. Similarly, in the 2024 Stack Overflow Developer Survey \citep{stackoverflow2024Stack}, which collected over 65,000 responses, 82\% of developers using AI tools reported using them to write code. The 2025 survey \citep{stackoverflow2025Stack}, based on over 49,000 responses, suggested that adoption is becoming more diversified across development tasks, with synthetic data generation, debugging, and other tasks gaining prominence over code writing. Nevertheless, developers resist using AI for high-responsibility tasks, such as deployment and monitoring \citep{stackoverflow2025Stack}.}

\textcolor{black}{The scarcity of PTMs for requirements engineering and software design may further be related to the inherent difficulty of automating and evaluating these activities. Requirements are often difficult to specify in AI-based systems \citep{habiba2024mature}, and current requirements engineering applications have been found to be insufficiently adaptable to the development of AI systems \citep{ahmad2023requirements}. In fact, a recent survey \citep{REFSQ2026} shows that the use of generative AI in requirements engineering is not commonplace yet, due mainly to ethical and legal concerns, as well as fear of hallucinations. When considering software architecture, many concerns arise, including LLM accuracy and challenges in evaluation metrics \citep{esposito2025generative}. Moreover, the software architecture community lacks benchmarks for evaluating generative AI capabilities on architectural tasks  \citep{adnan2026archbench}. These challenges make requirements and design tasks harder to evaluate than implementation-oriented tasks, for which there exists a large number of benchmarks \citep{abdollahi2025surveying}. Furthermore, current models are particularly well-suited to code generation and repair, which rely on textual artifacts, compared with more abstract SE activities such as requirements engineering and software design \citep{10.1145/3695988}.}

\subsection{Related Work}\label{sec:related_work} 
HF offers a rich ecosystem for conducting empirical studies. Recent repository mining studies have explored the HF ecosystem from several angles to assess its utility and characteristics. \citet{ait2025suitability} evaluated the suitability and potential of the platform for supporting empirical studies, and \cite{jiang2023empirical} addressed PTM reuse. Researchers have also investigated standardization and release practices within the ecosystem, including the study of PTM naming conventions \citep{jiang2025see}, and the analysis of release transparency, versioning, and naming \citep{ajibode2025towards}. Other works have focused on the lifecycle and management of these assets, such as \cite{castano2024machine}, who conducted a large-scale longitudinal study on how models change over time. While these studies characterize the broader HF landscape, our work specifically targets the identification and cataloguing of SE PTMs.

Key aspects of the related work are summarized in Table \ref{tab:related_work}. 
A systematic literature review conducted by \citet{10.1145/3695988} analyzed 395 research papers from January 2017 to January 2024 and categorized LLMs into SE tasks. These tasks were grouped into SE activities according to the SDLC. The study identified 88 distinct SE tasks, revealing that LLMs are most commonly applied in the software development activity, with \textit{code generation} and \textit{program repair} being the most prevalent tasks for employing LLMs in software development and maintenance activities, respectively. In contrast, the earlier SDLC phases, such as requirements engineering and software design, are the least explored in the context of LLM applications.

\begin{table}[!htb]
    \centering
    % \scriptsize
    \caption{Summary of relevant aspects in related work on cataloguing AI models for SE}
    \label{tab:related_work}
    \begin{adjustbox}{max width=\textwidth}
    \begin{tabular}{ccccccc}
        \toprule
        \textbf{Study} & \makecell{\textbf{Empirical} \\ \textbf{Standard} \citep{githubEmpiricalStandardsdocsstandardsMaster}} & \textbf{SDLC} &\textbf{Classification} &  \makecell{\textbf{Information} \\ \textbf{Used}} & \makecell{\textbf{Initial} \\ \textbf{Resources}} & \makecell{\textbf{Identified} \\ \textbf{Resources}}\\
        \midrule
         \citet{10.1145/3695988} & \makecell{Systematic \\ Review} & \cmark & \makecell{88 SE tasks \\ across \\ 6 SE activities} & Paper content & 218,765 papers & 395 papers\\
         \midrule
         \citet{10.1145/3661167.3661215} & \makecell{Repository \\ Mining} & \xmark & 6 macro-tasks & HF model card & 381,240 PTMs & \makecell{Tested on: \\  BERT, \\RoBERTa, \\T5}\\
         \midrule
         \citet{yang2024ecosystemlargelanguagemodels} & \makecell{Repository \\ Mining} & \xmark & 9 categories & \makecell{HF model \\ documentation} & 1,566 PTMs & 366 PTMs\\
         \midrule
         \citet{wang2025software} & \makecell{Systematic \\ Review} & \cmark & \makecell{30 SE tasks \\ across\\ 6 SE activities} & \makecell{Academic \\ publications} & 1,347 papers & \makecell{181 benchmarks \\ from \\ 461 relevant papers} \\
         \midrule
         \citet{ferino2025novice}  & \makecell{Systematic \\ Review}  & \cmark &  \makecell{35 SE tasks\\ across\\ 5 SE activities} & Paper content & 501 studies & 80 studies \\
         \midrule
         \citet{niu2022deep} & \makecell{Survey}  & \xmark & \makecell{18 tasks \\ across \\ understanding \\ and generation} & \makecell{Model \\technical \\details} & 20 PTMs & 20 PTMs \\ 
         
         \bottomrule
         \makecell{This \\ study} & \makecell{Repository \\ Mining} & \cmark & \makecell{74 SE tasks \\ (from a taxonomy \\ of 146 SE tasks) \\ across \\ 5 SE activities} & \makecell{HF model card \\ description, \\ metadata, \\ arXiv abstract} & \textcolor{black}{2,929,425} PTMs & \textcolor{black}{5,027} PTMs \\
         \bottomrule
    \end{tabular}
    \end{adjustbox}
\end{table}

Work by \citet{10.1145/3661167.3661215} highlighted the lack of an SE classification of PTMs in HF, as the existing one is specific for ML. To address this gap, they defined six macro-tasks and proposed a semi-automated approach to extract SE-relevant information from model cards. However, their validation was limited to just three PTMs (\textit{BERT}, \textit{RoBERTa}, and \textit{T5}), and their categorization did not align with the SDLC.

Similarly, \citet{yang2024ecosystemlargelanguagemodels} analyzed the ecosystem of LLMs as of August 2023, curating 366 PTMs from HF for SE. The study also explores the use of LLMs to assist in constructing and analyzing the ecosystem, which increased the model size by a factor of 16.5. They focus on code-based LLMs, which represent a more specific scope compared to LLMs for broader SE tasks. 

\citet{wang2025software} examined benchmarks for code LLMs and agents, revealing an imbalance in their distribution across the SDLC. Around 60\% of existing benchmarks focus on the software implementation phase, whereas requirements engineering and software design phases are underrepresented.

Complementary to these efforts, \citet{ferino2025novice} conducted a systematic literature review of 80 studies to investigate LLM adoption among novice developers. Their findings indicated that LLMs are being used across SDLC phases (except for software management), with most of the addressed tasks belonging to software development and little attention on software design. In their study, they organized the activities according to the SDLC, but grouped some of them together (i.e., requirements engineering and software design, software development and software quality assurance, software maintenance). 

\citet{niu2022deep} provided a survey of PTMs for code. They categorized 20 CodePTMs across 18 SE tasks and four dimensions: architecture, modality, training tasks, and programming languages. Their taxonomy emphasized the distinction between understanding and generation tasks, highlighting the expanding role of LLMs in automating software implementation activities. 

Beyond academic studies, community platforms such as \citet{paperswithcode} also offer structured connections between papers and code implementations, offering valuable categorizations of ML tasks \footnote{In July 2025, Papers With Code (PWC) was integrated into HF as \textit{Trending Papers} (https://huggingface.co/papers/trending). Although the original website is no longer available, the historical PWC data remains accessible via HF datasets (e.g., https://huggingface.co/datasets/pwc-archive/papers-with-abstracts).}. However, these are primarily oriented toward traditional ML domains. Even though the platform includes an SE-related category (i.e., \textit{computer code}), it combines general ML tasks (e.g., \textit{reinforcement learning}, \textit{semantic segmentation}) with SE-specific tasks (e.g., \textit{code generation}, \textit{program repair}), limiting its utility for SE-focused research and practice.

Despite the notable advances mentioned above, a comprehensive SE-centric classification framework of PTMs in HF remains absent. Existing efforts, such as those by \citet{10.1145/3695988} and \citet{wang2025software}, are grounded in static academic literature and are not designed to analyze or update classifications based on the contents of repositories. Other repository-based studies do not align their categorization with the SDLC or focus narrowly on a subset of code-centric LLMs \citep{10.1145/3661167.3661215,yang2024ecosystemlargelanguagemodels}. \citet{10.1145/3661167.3661215} take a step toward automation with a semi-automated mapping based on model cards, but their classification remains limited in scope and adaptability. In contrast, this paper adopts an SE perspective to extend the existing taxonomy in \citet{10.1145/3695988}, to encompass a broader and more diverse set of PTMs, cataloguing them according to specific SE activities and tasks within the SDLC. By analyzing a substantially larger and broader set of PTMs, this work addresses unique SE-specific requirements not met by previous studies. As a result, we propose a novel classification framework tailored to the SE community. 

\subsection{Comparison with prior studies}
Our results align with \citet{10.1145/3695988}, both identifying a strong focus on \textit{code generation} tasks, reflecting the broad adoption of generative PTMs in development workflows. As shown in our findings, the number of PTMs associated with software implementation is more than \textcolor{black}{nine} times higher than that of the next most represented SE activity, \textcolor{black}{software quality assurance}. Our study advances the field by analyzing a substantially larger and more diverse set of \textcolor{black}{5,027} PTMs, compared to the literature-based analysis by \citet{10.1145/3695988}.

Consistent with \citet{wang2025software}, who focus on code LLMs, we observe this skew toward software implementation. Yet, our work broadens the scope by covering a wider range of SE tasks and phases (identifying \textcolor{black}{74} distinct SE tasks across five SDLC activities), providing finer granularity and alignment with the full lifecycle.
This imbalance highlights the urgent need for greater research and tooling support for early SDLC activities, such as requirements engineering and software design. The growing interest in PTMs simulating diverse SE roles, such as MetaGPT \citep{hong2023metagpt}, highlights this gap.

\citet{ferino2025novice} focused on LLM adoption among novice developers, highlighting that the most addressed tasks belonged to the software \textcolor{black}{implementation} activity and software maintenance. At the same time, they also identified challenges (i.e., \textcolor{black}{loss of} learning opportunities) and suggested best practices (e.g., prompt engineering). Nevertheless, their study concentrates only on LLMs, with a focus on novice developers' practices. 

\begin{figure}[!htb]
    \centering
   \includegraphics[width=1\linewidth]{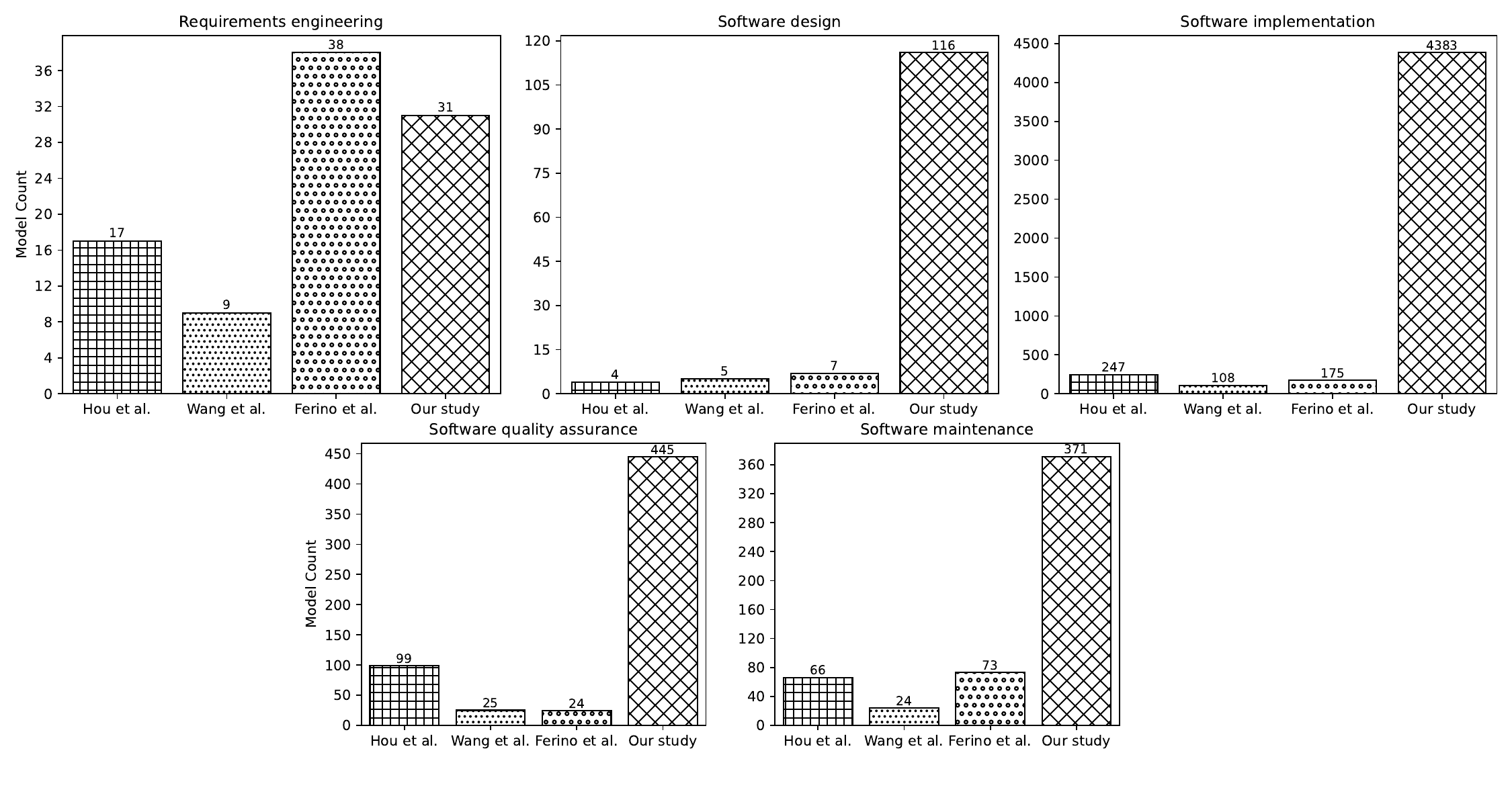}
    \caption{Distribution of models per SE activity among related works that are aligned to the SDLC}
    \label{fig:discussion_model_counts_per_se_activity}
\end{figure}

In terms of SDLC representation, Figure~\ref{fig:discussion_model_counts_per_se_activity} illustrates the number of models identified by each of the three aforementioned studies \citep{10.1145/3695988,wang2025software,ferino2025novice} and ours. 
All four studies show that software implementation is the dominant SE activity in terms of model coverage, a trend that is most pronounced in our study, with \textcolor{black}{4,383} PTMs. 

Regarding requirements engineering, we have catalogued a similar number of resources as \citet{ferino2025novice}. Notably, \citet{ferino2025novice} grouped multiple SE tasks into broader SE activities, which, for instance, included 20 studies employing LLMs for \textit{benchmarking} as part of the combined activity ``Requirements Engineering \& Software Design".

For software design, all previous studies analyzed a similar number of models (4, 5, and 7), while we found \textcolor{black}{116 PTMs in HF for such activity}, which may be explained by our broader scope and the inclusion of more recent PTMs. 

The overwhelming prevalence of software implementation PTMs (\textcolor{black}{4,383}) compared to early-stage activities warrants specific discussion. This disparity is likely driven by the abundance of structured code datasets and mature execution-based benchmarks which facilitate development. 

For the case of software quality assurance, our study identified a considerable jump, cataloguing \textcolor{black}{445} models, significantly surpassing the next highest count of 99 by \citet{10.1145/3695988}, and exceeding the counts of 25 and 24 from \citet{wang2025software} and \citet{ferino2025novice}, respectively. This suggests greater activity for software quality assurance that previous studies may have missed. 

Regarding software maintenance, our study again registered the highest count, with \textcolor{black}{371} models, compared to 66 for \citet{10.1145/3695988}, 24 for \citet{wang2025software}, and 73 for \citet{ferino2025novice}. This indicates that, while still far behind implementation, maintenance is a growing area for SE PTMs.

\subsection{Overlap with large HF datasets}
We have also compared how our catalogue overlaps with two large, generic HF-based datasets to contextualize its coverage. HFCommunity \citep{10123660} aggregates data from the HF Hub, while PeaTMOSS \citep{10.1145/3643991.3644907} comprises metadata for 281,638 PTMs and detailed snapshots for those with over 50 monthly downloads (14,296 PTMs), and links PTMs with 28,575 open-source software repositories from GitHub that use them. 

We found that \textcolor{black}{$981$ ($19.51\%$) and $244$ ($4.85\%$)} of our SE PTMs appear in HFCommunity and PeaTMOSS, respectively. This overlap ($<0.2\%$ of those collections) arises from the recency of our catalogue, as HFCommunity has been static since October 2024 and PeaTMOSS since August 2023, and from the inclusivity of our catalogue, which retains specialized and emerging PTMs regardless of their popularity. 

Table \ref{tab:peatmoss_community} quantifies this temporal gap by illustrating the number of PTMs in our study released after the cutoff dates of the comparison datasets. As both have been static for a long period, they miss a significant portion of the recent surge in SE models. For instance, we identified \textcolor{black}{4,151} software implementation models and \textcolor{black}{437} quality assurance models that were published after the PeaTMOSS snapshot. Even compared to the more recent HFCommunity, our catalogue captures \textcolor{black}{3,063} additional implementation models released after October 2024. This demonstrates the rapid evolution of the ecosystem. However, these partial intersections create opportunities for richer analysis, since combining our catalogue with broader datasets can reveal ecosystem-level insights that remain hidden when these resources are considered in isolation.

\begin{table}[!h]
    \centering
    \caption{Number of SE PTMs in our catalogue published after the data collection cutoff dates of PeaTMOSS and HFCommunity}
    \begin{tabular}{ccc}
        \toprule
        \textbf{SE Activity} & \makecell{\textbf{After PeaTMOSS }\\ (Aug 2023)} & \makecell{\textbf{After HFCommunity} \\(Oct 2024)} \\
        \midrule
        Requirements engineering & \textcolor{black}{27} & \textcolor{black}{22}\\
        \midrule
        Software design & \textcolor{black}{115} & \textcolor{black}{105}\\
        \midrule
        Software implementation & \textcolor{black}{4,151} & \textcolor{black}{3,063}\\
        \midrule
        Software maintenance & \textcolor{black}{345} & \textcolor{black}{281}\\
        \midrule
        Software quality assurance & \textcolor{black}{437} & \textcolor{black}{370}\\
        \bottomrule
    \end{tabular}
    \label{tab:peatmoss_community}
\end{table}

\subsection{Implications}
The findings of this study \textcolor{black}{have practical implications} for researchers and practitioners seeking to navigate the rapidly growing landscape of SE PTMs. \textcolor{black}{Prior empirical research has shown that PTM reuse is not a straightforward process, as many challenges may arise, including missing attributes, discrepancies between claims and actual performance, and model risks \citep{jiang2023empirical}. These obstacles are increasingly relevant as AI tools are regularly being used by developers for coding and development tasks \citep{jetbrainsWhichCoding}. Moreover, they use multiple LLMs in their work, including OpenAI GPT, Gemini, and DeepSeek, highlighting the importance of model selection \citep{stackoverflow2025Stack}.}

By aligning open-source PTMs in HF with the SDLC, our catalogue supports informed \textcolor{black}{reuse of PTMs in software projects}. 
\textcolor{black}{To illustrate one potential use case,} consider an SE team at a US-based company developing a code completion tool for Python, integrated into their existing desktop IDE. This team aims to improve developer productivity through automation while prioritizing environmental sustainability and compliance with open-source licensing policies.
Our catalogue supports selecting SE PTMs that best align with specific needs, thereby streamlining decision-making processes \citep{ben2025should}. As characterized by \citet{zhouunifying}, a combination of semantic and metric-based selection criteria can be set. \textcolor{black}{Using our catalogue, the team can first identify PTMs associated with software implementation} and refine the results to those explicitly supporting Python. Then, they can restrict the selection to PTMs reporting energy consumption, aligning with their sustainable AI practices. From this subset, they identify PTMs released under permissive licenses, such as MIT. When multiple candidates meet all criteria, the team evaluates them based on quality attributes such as accuracy and benchmark performance. Through this process, the team identifies \texttt{ashwinR/CodeExplainer} \citep{huggingfaceAshwinRCodeExplainerHugging}. \textcolor{black}{This example demonstrates how the catalogue} can support decision-making \textcolor{black}{when selecting SE PTMs} in real-world scenarios.
%Figure \ref{fig:selection_diagram} depicts this selection process, emphasizing how our approach guides users to find PTMs best suited to their specific context. 

% \begin{figure}[!h]
%     \centering
%     \includegraphics[width=0.6\linewidth]{selection_diagram.pdf}
%     \caption{Selection diagram}
%     \label{fig:selection_diagram}
% \end{figure}

\begin{figure}[!h]
    \centering
    \includegraphics[width=1\linewidth]{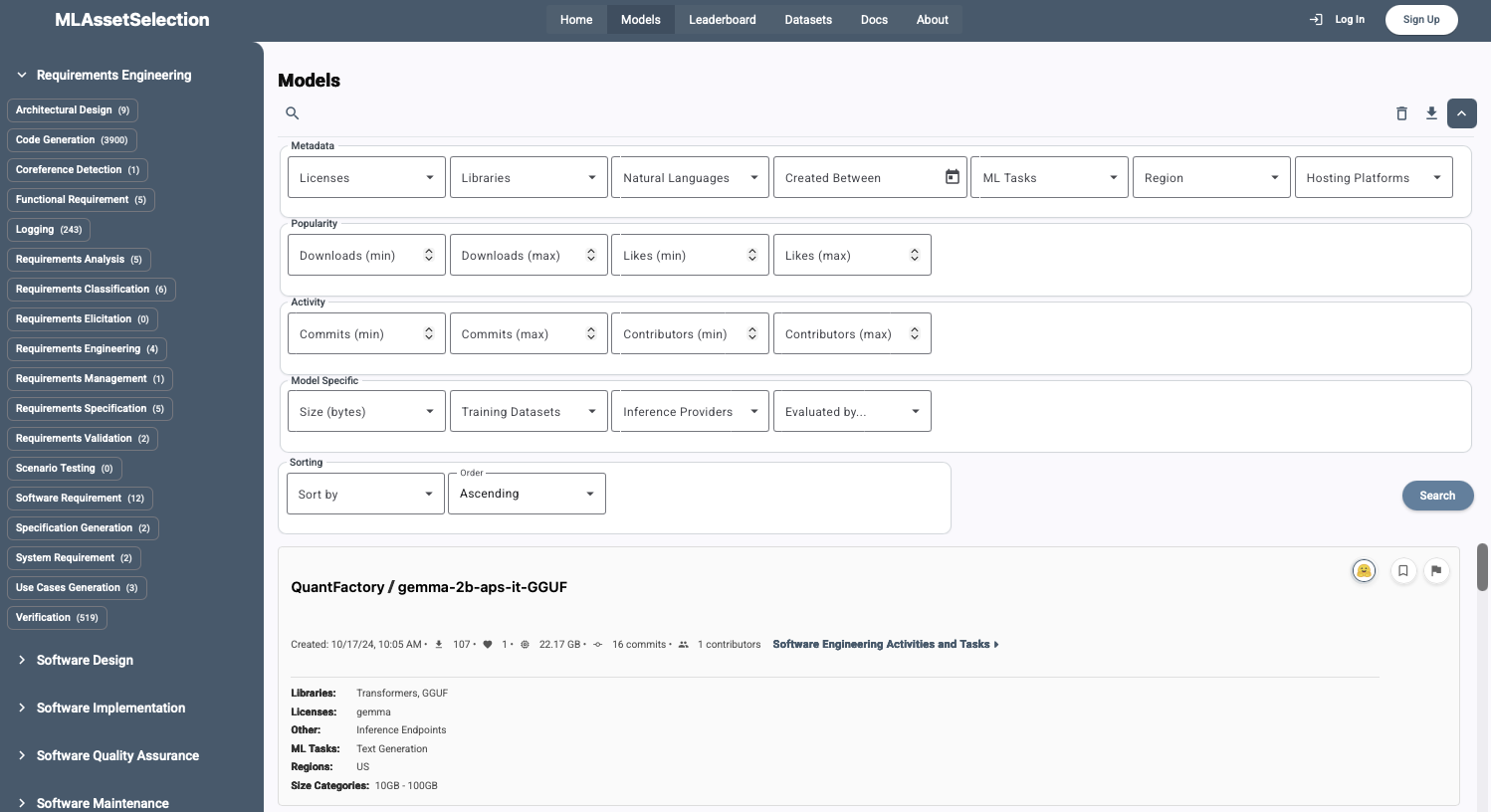}
    \caption{MLAssetSelection overview}
    \label{fig:MLAssetSelection}
\end{figure}

The findings of this study have motivated the ongoing development of \textit{MLAssetSelection}, a web application that operationalizes our cataloguing pipeline to assist in the sampling and selection of SE PTMs \citep{gonzalez2026tool}. As illustrated in Figure~\ref{fig:MLAssetSelection}, the tool provides an interactive interface for exploring the catalogue. Users can filter PTMs by \textit{identifier}, \textit{scope} (e.g., SE activities and tasks, ML tasks), \textit{metadata} (e.g., licenses, libraries, natural languages, creation date, size categories: bytes for models and rows for datasets), \textit{popularity} (e.g., downloads, likes), \textit{activity} (e.g., commits, contributors), and \textit{model-specific attributes} (e.g., training dataset, region, inference providers, evaluation). 

From a research perspective, our findings reveal important gaps and opportunities in the current SE PTM ecosystem. Certain phases of the SDLC remain underrepresented, particularly requirements engineering and software design, where few PTMs explicitly target these activities. Beyond identifying such gaps, our catalogue provides a foundation for researchers to delve deeper into the challenges uncovered in this study and to conduct empirical studies using the collected SE PTMs.

\subsection{Updatability}
The replication package \citep{replication_package} accompanying this work includes the complete pipeline described in Section \ref{sec:methodology}, enabling full replicability of our results. By following the documented steps, users can reproduce the same cataloguing of PTMs available up to \textcolor{black}{May 31, 2026}, while also retrieving and classifying PTMs added since that date. This supports replicability and allows others to validate or extend our work. 

Given the exponential growth of PTMs in HF \citep{ait2025suitability,10.1145/3643991.3644898}, re-executing the full pipeline for the entire repository is computationally inefficient and arguably unnecessary unless the taxonomy itself requires modification (e.g., updating the taxonomy of SE tasks or validating the classification strategy). To maintain an up-to-date dataset, our framework supports incremental updates: users can retrieve only the PTMs created after the cutoff date, catalogue them, and merge the results with our baseline dataset. This ensures efficient maintainability without \textcolor{black}{the} redundant \textcolor{black}{carbon footprint of a full re-execution.}

We acknowledge that a research paper is inherently static (our \textcolor{black}{detailed} analysis captures the state of HF as of \textcolor{black}{May 31, 2026}), whereas the HF ecosystem is highly dynamic. \textcolor{black}{The number of repositories has been growing at a rapid pace \citep{laufer2025anatomy}, and has continued to grow since our data collection.} Given that \textcolor{black}{thousands} of new repositories are created daily \citep{huggingfaceclemHugging}, the absolute numbers presented in this article will inevitably become outdated (as in any other article of this kind). To address this temporal limitation, the value of this work lies not only in the static snapshot provided but also in the updatability of the framework. While the raw numbers presented in this article reflect the state of HF as of \textcolor{black}{May 31, 2026}, we argue that the distribution of SE tasks and \textcolor{black}{the proposed} taxonomy \textcolor{black}{provide a solid baseline for understanding the current landscape. Nevertheless, we acknowledge that in such a fast-moving field, periodic updates to the taxonomy will be necessary to capture emerging tasks.} By applying our validated filtering and cataloguing methodology to new entries, the community can maintain a living catalogue. This ensures that the insights regarding the distribution of SE tasks and the utility of the taxonomy remain relevant, even as the underlying raw numbers of the HF ecosystem continue to rise. 

Furthermore, our catalogue is designed to be extensible and interoperable with other existing datasets. Although our work focuses specifically on SE, it can be integrated with broader, generic catalogues such as HFCommunity \citep{10123660} or PeaTMOSS \citep{10.1145/3643991.3644907}.

\section{Threats to Validity}\label{sec:threats}
Potential threats that could impact our study's validity are outlined. 
\begin{itemize}
    \item \textbf{Internal Validity}: The quality of model card descriptions may pose a threat, as incomplete or ambiguous documentation could lead to misclassification. To mitigate this, we enriched our data collection process by incorporating additional sources of evidence, such as metadata fields and abstracts from linked arXiv publications, thereby improving the robustness of our classification decisions. 

    \item \textbf{External Validity}: While HF is a leading platform for sharing PTMs, other registries, such as PyTorch Hub \citep{pytorchPyTorch}, may host relevant PTMs that are not captured in our analysis. Moreover, ongoing changes in HF’s platform structure, documentation practices, or model metadata could affect the broader applicability of our conclusions. 
    
    \item \textbf{Construct Validity}: Our catalogue of SE tasks and activities builds upon an established taxonomy derived from existing literature \citep{10.1145/3695988}. We extended this taxonomy using SWEBOK \textcolor{black}{\citep{washizaki2024swebok} and iterative discussions among the authors to include a broader set of 146 SE tasks. However, our taxonomy was not externally validated by independent SE experts or practitioners and may evolve as the SE community introduces new standards and practices. Additionally, our pipeline depends on lexical matching, which risks missing PTMs that employ divergent phrasing. We mitigated this threat by extracting data across three distinct sources: model card descriptions, metadata, and arXiv abstracts.}
    
    \item \textbf{Conclusion Validity}: While manual annotation may introduce subjective bias, we mitigated this threat through inter-coder agreement practices. We acknowledge the use of Gemini to support part of the classification process. To mitigate potential risks, we designed \textcolor{black}{a two-phase validation protocol. First, we conducted five pilot tests} (one per SE activity), and validated outputs on a statistically significant sample. A total of 1,346 samples were manually reviewed across all five pilots, \textcolor{black}{with a 10\% subset of each pilot independently annotated by multiple authors to assess ICR. Second, we evaluated the generalization of the resulting prompts on a separate subset of samples not used during the pilot phase. Overall, these measures strengthen the robustness of the process, although the approach remains grounded in Gemini, which may introduce model-specific variability.}
\end{itemize}

To further address these threats, we have made our entire study fully replicable. All data, scripts, and resources are publicly available, allowing future researchers to validate, challenge, or build upon our findings as the ecosystem of SE PTMs continues to evolve.

\section{Conclusion and Future Work}\label{sec:conclusion}
This study examines the availability of SE PTMs hosted on HF, identifying \textcolor{black}{5,027} PTMs relevant to SE. Software design and requirements engineering are under-represented, highlighting a need for broader coverage across the SDLC. The most prevalent ML task across SE-related PTMs is \textit{text generation}, while \textit{code generation} is the most commonly addressed SE task. Additionally, there has been a significant surge in SE PTMs since 2023 Q2. This snapshot of the current landscape highlights the existing gaps in SE resources and provides valuable insights into the field's evolving needs, helping to motivate future research efforts. By classifying PTMs according to SE activities, our work also contributes to automating the sampling and selection of relevant PTMs for researchers and practitioners.

For future work, we plan to extend this classification to other registries, such as \textcolor{black}{Kaggle \citep{KaggleModels}} and GitHub Models \citep{githubGitHubModels}, allowing for a broader analysis of SE PTMs. We also aim to enhance the catalogue by recovering and integrating relevant information from original base model cards, which may not be explicitly stated in derivative model cards. 
%To make our classification more actionable, we propose developing a dashboard that enables automatic sampling and selection of PTMs for real-world applications, supporting both researchers in conducting empirical studies and practitioners in integrating PTMs into software engineering workflows. 
\textcolor{black}{In addition, we will investigate methodological improvements to strengthen the robustness of the classification process, including the use of multiple LLMs beyond a single-model setup and external validation with SE experts and practitioners to further assess the proposed taxonomy.}
We further intend to explore how this classification can be embedded into SE pipelines by enabling automatic recommendation and adaptation of PTMs to SE tasks, such as \textit{design pattern} or \textit{bug detection}. 
\textcolor{black}{More broadly, we encourage the SE community to develop standardized, community-accepted benchmarks for long-tail SE activities, particularly requirements engineering and software design, where the evaluation evidence we observed is both scarce and heterogeneous.}

% \section*{Acknowledgment}

\section*{Declarations}

\bmhead{Author contributions}
\begin{itemize}
    \item Alexandra González: Conceptualization, Methodology, Software, Validation, Formal Analysis, Investigation, Data Curation, Writing - Original Draft, Visualization.
    \item Xavier Franch: Conceptualization, Methodology, Supervision, Validation, Writing - Review \& Editing 
    \item David Lo: Writing - Review \& Editing
    \item Silverio Martínez-Fernández: Conceptualization, Methodology, Supervision, Validation, Writing - Review \& Editing
\end{itemize}

\bmhead{Data availability statement}
To support transparency, reproducibility, and replicability, and in line with the EMSE open-science guidelines \footnote{\url{https://github.com/emsejournal/openscience}}, we provide a publicly available replication package containing all research artifacts used in this study \citep{replication_package}. It includes the taxonomy of 146 SE tasks aligned with the SDLC, along with raw and preprocessed data, and the complete pipeline for data collection, preparation, search, and filtering. We also share the manual annotations made by human experts during the study.

\bmhead{Funding}
This work was supported by the FI-STEP predoctoral fellowship (grant no.\ 2025 STEP 00407) awarded by AGAUR (Government of Catalonia) and co-funded by the European Social Fund Plus (FSE+) of Catalonia 2021--2027, and by Grant PID2024-156019OB-I00 funded by MICIU/AEI/10.13039/ 501100011033 and by ERDF, EU. 

\bmhead{Conflict of interests}
The authors declare that they have no conflict of interest.

\bmhead{Ethical approval}
This study does not involve human participants or animals.

\bmhead{Informed consent}
No human subjects were involved in this study.

\bmhead{Clinical Trial Number}
Not applicable.

\bibliographystyle{spbasic}
\bibliography{references}

\newpage
\appendix
\section{SE Activities and Task Taxonomy}\label{sec:se_tasks} 
\textcolor{black}{Table \ref{tab:se_tasks} presents the complete taxonomy of SE activities and their corresponding SE tasks, extending the original classification proposed in \cite{10.1145/3695988}. For detailed information on how it was constructed, please refer to Section \ref{subsubsec:task_identification}.}

\begin{longtable}{p{4cm} p{5cm} r}
\caption{\textcolor{black}{Taxonomy of SE tasks per activity, with the number of SE PTMs matched per task (May 2026)}}
\label{tab:se_tasks} \\

\toprule
\textbf{SE Activity} & \textbf{SE Tasks} & \textbf{Count} \\
\midrule
\endfirsthead

\toprule
\textbf{SE Activity} & \textbf{SE Tasks} & \textbf{Count} \\
\midrule
\endhead

\bottomrule
\endfoot

\bottomrule
\endlastfoot

Requirements engineering & software requirement & 11 \\
 & requirements classification & 6 \\
 & requirements analysis & 4 \\
 & requirements engineering & 4 \\
 & requirements specification & 4 \\
 & functional requirement & 3 \\
 & use cases generation & 3 \\
 & requirements validation & 2 \\
 & specification generation & 2 \\
 & system requirement & 2 \\
 & coreference detection & 1 \\
 & anaphoric ambiguity treatment & 0 \\
 & nonfunctional requirement & 0 \\
 & quality of service constraints & 0 \\
 & requirement analysis and evaluation & 0 \\
 & requirements change control & 0 \\
 & requirements elicitation & 0 \\
 & requirements management & 0 \\
 & requirements review & 0 \\
 & requirements scrubbing & 0 \\
 & software product requirement & 0 \\
 & software project requirement & 0 \\
 & specification formalization & 0 \\
 & technology constraints & 0 \\
 & traceability automation & 0 \\

\midrule

Software design & system design & 64 \\
 & design pattern & 24 \\
 & rapid prototyping & 16 \\
 & software architecture & 12 \\
 & architectural design & 6 \\
 & architecture pattern & 5 \\
 & software design & 5 \\
 & architectural decisions & 2 \\
 & structural design & 2 \\
 & architecture framework & 1 \\
 & architecture style & 0 \\
 & behavioural design & 0 \\
 & design rationale & 0 \\
 & detailed design & 0 \\
 & gui retrieval & 0 \\
 & high-level design & 0 \\
 & model-based design & 0 \\
 & software specification synthesis & 0 \\

\midrule

Software implementation & code generation & 3,100 \\
 & coding & 1,278 \\
 & code completion & 572 \\
 & data analysis & 162 \\
 & code search & 119 \\
 & code understanding & 110 \\
 & program synthesis & 84 \\
 & code translation & 83 \\
 & code editing & 50 \\
 & code summarization & 48 \\
 & code representation & 23 \\
 & instruction generation & 13 \\
 & api recommendation & 6 \\
 & code comment generation & 6 \\
 & type inference & 6 \\
 & code recommendation & 2 \\
 & construction integration & 1 \\
 & software implementation & 1 \\
 & api documentation augment & 0 \\
 & api documentation smells & 0 \\
 & api entity and relation extraction & 0 \\
 & api inference & 0 \\
 & construction quality & 0 \\
 & control flow graph generation & 0 \\
 & fuzz driver generation & 0 \\
 & identifier normalization & 0 \\
 & method name generation & 0 \\
 & reuse in construction & 0 \\

\midrule

Software quality assurance & verification & 151 \\
 & information security & 93 \\
 & vulnerability detection & 67 \\
 & cybersecurity & 42 \\
 & static analysis & 35 \\
 & construction for security & 28 \\
 & defect detection & 23 \\
 & test generation & 21 \\
 & decompilation & 13 \\
 & quality assurance & 11 \\
 & security testing & 8 \\
 & software testing & 6 \\
 & testing automation & 6 \\
 & software quality & 5 \\
 & bug localization & 4 \\
 & software security & 4 \\
 & gui testing & 3 \\
 & exploratory testing & 2 \\
 & security design & 2 \\
 & quality management & 1 \\
 & vulnerability management & 1 \\
 & binary taint analysis & 0 \\
 & compiler fuzzing & 0 \\
 & control flow testing & 0 \\
 & data flow testing & 0 \\
 & equivalence partitioning & 0 \\
 & error guessing & 0 \\
 & evidence-based testing & 0 \\
 & fault localization & 0 \\
 & invariant prediction & 0 \\
 & malicious code localization & 0 \\
 & mobile app crash detection & 0 \\
 & resource leak detection & 0 \\
 & scenario testing & 0 \\
 & security patterns & 0 \\
 & security requirements & 0 \\
 & state transition testing & 0 \\
 & syntax testing & 0 \\
 & test prediction & 0 \\

\midrule

Software maintenance & code review & 134 \\
 & logging & 117 \\
 & debugging & 70 \\
 & sentiment analysis & 49 \\
 & program repair & 15 \\
 & software maintenance & 9 \\
 & code clone detection & 6 \\
 & corrective maintenance & 6 \\
 & log parsing & 2 \\
 & bug prediction & 1 \\
 & bug reproduction & 1 \\
 & code coverage prediction & 1 \\
 & duplicate bug report detection & 1 \\
 & preventive maintenance & 1 \\
 & adaptive maintenance & 0 \\
 & additive maintenance & 0 \\
 & api misuses repair & 0 \\
 & bug triage & 0 \\
 & code review explained & 0 \\
 & code revision & 0 \\
 & code-review defects repair & 0 \\
 & crash bug repair & 0 \\
 & dockerfile repair & 0 \\
 & emergency maintenance & 0 \\
 & patch correctness prediction & 0 \\
 & patch detection & 0 \\
 & perfective maintenance & 0 \\
 & program merge conflicts repair & 0 \\
 & rename refactoring & 0 \\
 & review/commit/code classification & 0 \\
 & tag recommendation & 0 \\
 & technical debt payback & 0 \\
 & traceability recovery & 0 \\
 & type error repair & 0 \\
 & vulnerability repair & 0 \\
 & web test repair & 0 \\
\end{longtable}

\end{document}